\documentclass[journal=aamick,manuscript=article]{achemso}
\usepackage[utf8]{inputenc}
\usepackage{graphicx}
\usepackage{chemfig}
\usepackage{siunitx}
\usepackage{dcolumn}
\usepackage{bm}
\usepackage{booktabs}
\usepackage[version=4]{mhchem}
\usepackage{amssymb}
\usepackage{array}
\usepackage{multirow}
\usepackage{placeins}
\usepackage{csquotes}
\usepackage{caption}
\usepackage{xcolor}

\DeclareSIUnit{\atom}{atom}
\DeclareSIUnit{\elementarycharge}{\text{\ensuremath{e}}}

\usepackage{float}

\newcommand{\beginsupplement}{%
  \setcounter{equation}{0}%
  \setcounter{figure}{0}%
  \setcounter{table}{0}%
  \renewcommand{\theequation}{S\arabic{equation}}%
  \renewcommand{\thefigure}{S\arabic{figure}}%
  \renewcommand{\thetable}{S\arabic{table}}%
  \providecommand{\theHequation}{}%
  \providecommand{\theHfigure}{}%
  \providecommand{\theHtable}{}%
  \renewcommand{\theHequation}{S\arabic{equation}}%
  \renewcommand{\theHfigure}{S\arabic{figure}}%
  \renewcommand{\theHtable}{S\arabic{table}}%
}

\title{Oxygen-deficiency-driven phase segregation enables enhanced hole transport in amorphous tellurium oxides}

\author{Rafael Costa-Amaral}
\email{costa.amaral.rafael@gmail.com}
\affiliation{Institute for Materials Research, Tohoku University, Sendai 980-8577, Japan.}

\author{Yu Kumagai}
\email{yukumagai@tohoku.ac.jp}
\affiliation{Institute for Materials Research, Tohoku University, Sendai 980-8577, Japan.}
\affiliation{Organization for Advanced Studies, Tohoku University, Sendai 980-8577, Japan}

\date{\today} 

\begin{document}

\begin{abstract}
Amorphous oxide semiconductors allow scalable electronics, yet high-mobility p-type
counterparts remain rare because O-$2p$ valence bands are typically deep and
spatially localized.
Motivated by recent reports of unusually
high hole mobilities in oxygen-deficient \ce{Se}-doped amorphous tellurium oxides ($a$-\ce{TeO$_x$}),
we investigated $a$-\ce{TeO$_x$} with and without \ce{Se} doping using machine-learning-accelerated ab initio molecular dynamics with
hybrid-functional defect calculations.
We find that oxygen depletion drives nanoscale segregation into interpenetrating $a$-\ce{Te} and $a$-\ce{TeO2} domains
with distinct roles: \ce{Te} vacancies in oxide-like/interfacial environments supply holes,
while transport is mediated by percolating \ce{Te}-$5p$ pathways within the $a$-\ce{Te} subnetwork.
Upon doping, we theoretically verify that \ce{Se} preferentially incorporates into the $a$-\ce{Te} domains enhancing connectivity.
This preference is nontrivial without explicit modeling given that \ce{Se} shares similar electronic
structure with both \ce{Te} and \ce{O}.
We further find that reducing the oxygen content can likewise enhance hole conductivity.
Finally, using amorphous \ce{SeO$_x$}, we show that domain segregation persists in
other amorphous chalcogen oxides, suggesting a transferable route to
achieving higher-mobility p-type amorphous oxides.
\end{abstract}

\maketitle
\section{Introduction}\label{sec1}

\begin{figure}[t!]
\centering\includegraphics[width=0.5\linewidth]{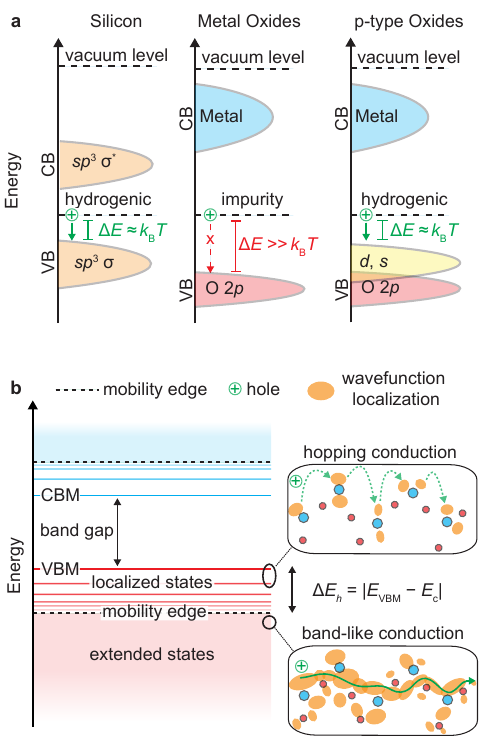}
\caption{(a) Schematic band alignment in crystalline \ce{Si}, conventional oxides, and p-type oxides.
In \ce{Si}, covalent $sp^3$ bonding places the VB near the vacuum level,
so acceptors form hydrogenic shallow states ($\Delta E \approx k_\mathrm{B}T$) and readily generate
mobile holes. In typical metal oxides, deep \ce{O}-$2p$-derived VBs yield large $\Delta E (\gg k_\mathrm{B}T$),
leading to localized holes and poor hole dopability. Hybridization of \ce{O}-$2p$ with cation $d^{10}$ or
$s^2$ lone pairs raises and delocalizes the VBM, enabling easier hole formation and band-like transport.
(b) Typical hole transport in an amorphous semiconductor. Orange clouds depict the spatial
extent of hole wavefunctions and green symbols indicate holes.
States near the VBM are localized and support thermally activated
hopping, whereas states below the mobility edge form extended bands
that enable band-like conduction.
The hole mobility-edge offset, $\Delta E_h = \lvert E_\mathrm{VBM} - E_c \rvert$, where
$E_c$ is the valence mobility-edge energy, measures the separation between
the VBM and extended states.}
\label{fig:band_scheme_amorphous}
\end{figure}

Amorphous oxide semiconductors are attractive for next-generation
complementary metal-oxide semiconductor (CMOS) and large-area electronics
because they can be deposited uniformly at low temperatures using scalable,
cost-efficient methods~\cite{Medvedeva_1700082_2017,Kim_2204663_2023}. While
n-type materials such as amorphous \ce{InGaZnO} ($a$-IGZO) are already
commercially successful~\cite{Kamiya_15_2010,Zhu_031101_2021}, high-mobility p-type amorphous oxides
remain elusive. As illustrated in Figure~\ref{fig:band_scheme_amorphous}a, the
valence-band maximum (VBM) of most oxides is dominated by deep \ce{O}-$2p$ states,
unlike crystalline \ce{Si} whose VBM composed of \textit{sp}$^3$ bonding lies much closer
to the vacuum level. This alignment makes acceptor doping intrinsically difficult
because dopants typically create deep localized hole states that are hard to thermally ionize at room temperature.
Hybridization with cation $d^{10}$ or $s^2$ lone-pair orbitals can
raise and delocalize the VBM, enabling p-type compounds~\cite{Vu_9505_2025,Kiyohara_246101_2025,
Kumagai_034063_2023,Hautier_2292_2013,Hu_140902_2020,Willis_11995_2021} such as \ce{Cu2O}~\cite{Mizuno_C179_2005},
\ce{SnO}~\cite{Varley_082118_2013}, and \ce{BaBiO3}~\cite{Schoop_5479_2013}.
In amorphous oxides, however, disorder broadens the upper valence band into localized
tail states and introduces a mobility edge (Figure~\ref{fig:band_scheme_amorphous}b), so carriers
must be thermally excited to reach extended states for band-like conduction.

Recently, \ce{Se}-doped $a$-\ce{TeO$_x$} thin-film transistors (TFTs) exhibit hole mobilities exceeding
\SI{20}{\square\cm\per\volt\per\second}~\cite{Liu_798_2024,Yang_e18364_2025}, with low-temperature transport
interpreted as band-like.
This behavior is intriguing given that crystalline tellurium oxides
are predicted to be wide-gap, difficult-to-dope insulators~\cite{Huyen_044065_2024,CostaAmaral_1605_2024,Xiao_016103_2025}.
Furthermore, the films are substantially oxygen deficient, suggesting a high density of oxygen vacancies that
are typically deep donors and would suppress p-type conduction~\cite{Oba_060101_2018}.
Based on spectroscopic evidence of elemental \ce{Te}, Liu et al.~\cite{Liu_798_2024}
proposed a \ce{Te}/\ce{TeO$_x$} mixed-phase interpretation, in which hole transport was suggested to proceed
through \ce{Te}-$5p$ states.
Additionally, the \ce{Se} doping enhanced the mobility, which was attributed to improved atomic network connectivity
from EXAFS signatures of \ce{Te-Se} bonding~\cite{Liu_798_2024}.
Nevertheless,
several key questions remain open:
(1) How does oxygen deficiency reorganize the amorphous network?
(2) Which microscopic states supply the hole carriers?
(3) How does \ce{Se} modify network connectivity to enable band-like hole transport?
In particular, questions (2) and (3) have not been clearly addressed with solid evidence from
either experiments or theoretical calculations.

Here, we address these questions by combining machine-learning-accelerated ab initio molecular dynamics
(AIMD) to generate amorphous networks across $x=0.0$--$2.0$ (including the experimentally
relevant $x=1.2$ composition~\cite{Liu_798_2024}) with hybrid-functional defect calculations.
To rationalize transport trends, we use the hole mobility-edge offset, ${\Delta}E_h$
(Figure~\ref{fig:band_scheme_amorphous}b), defined as the energy separation between the VBM and the
hole mobility edge, as a semi-quantitative proxy for the energetic accessibility of extended valence states, rather than as a direct mobility or conductivity.
In addition, we propose that analogous behavior with Se doping can be realized by controlling the oxygen content,
and discuss the extent to which the emergence of the hole conductivity is generalizable.

\section{Results and Discussion}\label{sec:results}

\subsection{Modeling oxygen-deficient amorphous \ce{TeO$_x$}}\label{sec:amorphous_benchmark}

\begin{figure}[t!]
\centering\includegraphics[width=0.5\linewidth]{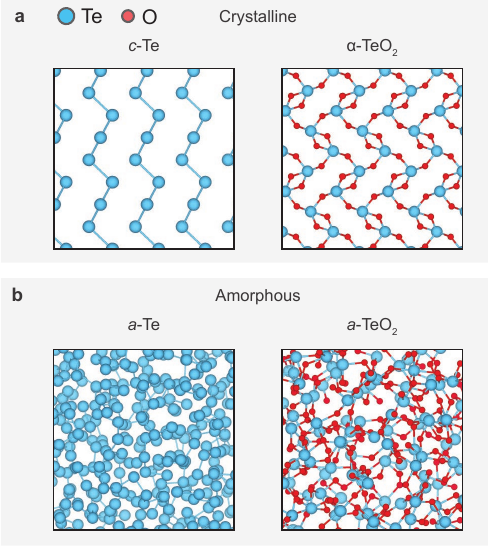}
\caption{
Atomic structures of (a) trigonal tellurium ($c$-\ce{Te}) and $\alpha$-\ce{TeO2}
    and (b) amorphous counterparts obtained by melt-quench AIMD simulations (see text for details).}
\label{fig:te_oxides}
\end{figure}

Figure~\ref{fig:te_oxides}a summarizes the characteristic bonding motifs of elemental Te and $\alpha$-\ce{TeO2}.
Elemental \ce{Te} favors twofold coordination and crystallizes as helical chains~\cite{HumeRothery_65_1930},
where relatively weak interchain interactions coexist with high hole mobilities~\cite{Champness_3038_1970}.
This chain motif arises from the two lone pairs on each \ce{Te} site.
By contrast, in the \ce{TeO2} polymorph, $\alpha$-\ce{TeO2}~\cite{Deringer_871_2014},
four-coordinated \ce{Te^{4+}} forms a three-dimensional network of distorted trigonal-bipyramidal units,
often described as \ce{TeO4} polyhedra with a lone-pair per \ce{Te} site oriented toward a vacant site.

\begin{figure*}[t!]
\centering\includegraphics[width=0.99\linewidth]{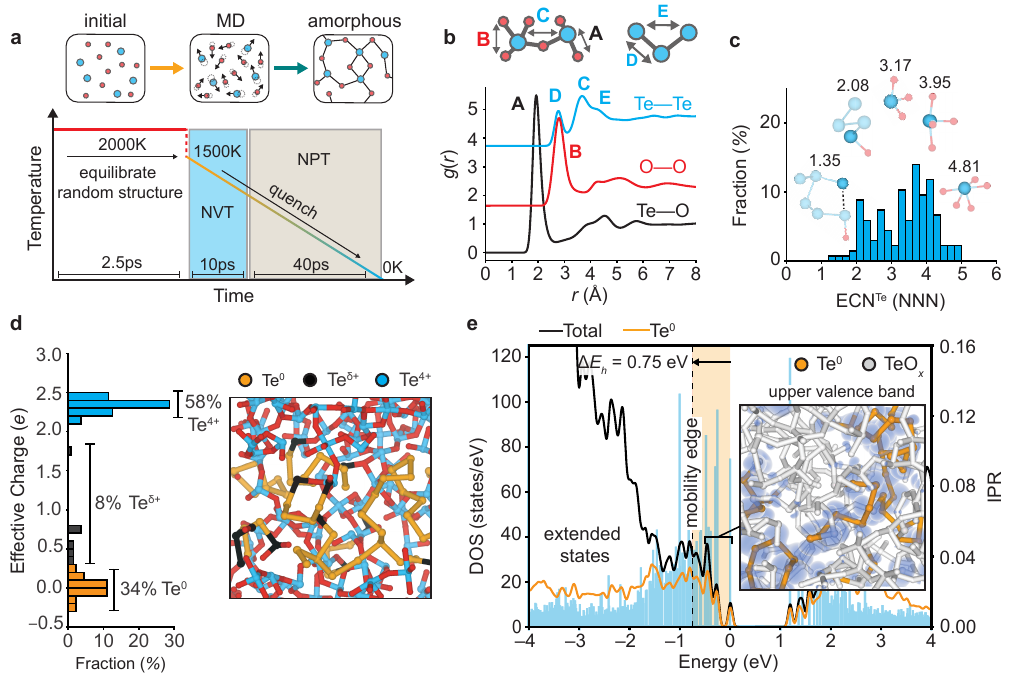}
\caption{Atomic and electronic structures of $a$-\ce{TeO$_{1.2}$}.
(a) Schematic of the melt-quench AIMD workflow used in this study. An initial
\SI{2000}{\kelvin} pre-equilibration precedes the \SI{1500}{\kelvin} quench shown schematically.
(b) Partial pair-correlation functions, $g(r)$ (\ce{Te-O}, \ce{O-O}, and \ce{Te-Te}),
    which are vertically shifted for comparison.
    Labeled peaks correspond to the local motifs in the insets.
(c) Histogram of the Te effective coordination number, $\mathrm{ECN}^{\ce{Te}}$, in number of nearest neighbors (NNN),
with representative fragments shown in the inset.
(d) Bader charge distribution revealing neutral \ce{Te^0}, partially oxidized \ce{Te$^{\delta+}$}, and \ce{Te$^{4+}$},
together with a structural snapshot highlighting the spatial
organization of \ce{Te$^0$}-rich regions (in orange).
(e) Total density of states (DOS) (black line), neutral-\ce{Te$^
0$}-projected DOS (orange line), and inverse participation ratio (IPR) (blue bars)
for $a$-\ce{TeO$_{1.2}$}. Energies are referenced to the VBM; the DOS is plotted on the
left axis and the IPR on the right axis. Localized tail states and the mobility edge are indicated.
The hole mobility-edge offset
$\Delta E_h$ is \SI{0.75}{\eV}. The inset maps \ce{Te$^0$} atoms and the charge-density isosurface of states within \SI{0.5}{\eV}
below the VBM, plotted at an isovalue of \SI{5}{\percent} of the maximum charge density.}
\label{fig:a_teox}
\end{figure*}

The amorphous structures were generated by quenching pre-equilibrated, randomly initialized configurations
using machine-learning-accelerated AIMD (Figure~\ref{fig:a_teox}a) and then relaxed at \SI{0}{\kelvin} with
a hybrid functional (see Methods for details). Unless otherwise stated, reported averages are obtained from
three independently initialized 300-atom amorphous cells per composition; sample-wise structural and electronic
quantities are provided in the Supporting Information (SI).
Because experimental benchmarks
for nonstoichiometric compositions are limited, we first validated the protocol against the
amorphous tellurium ($a$-\ce{Te}) and stoichiometric amorphous \ce{TeO2}
 ($a$-\ce{TeO2}), as shown in Figure~\ref{fig:te_oxides}b.
Accordingly, $a$-\ce{TeO2} can be viewed as a disordered network
that largely preserves the local \ce{TeO4} tetrahedral motifs.
Pair-correlation functions $g(r)$ and the effective coordination number\cite{DaSilva_023502_2011}
(ECN; Supporting Note~\num{3}) agree well with available experiments~\cite{Ichikawa_707_1973,Alderman_427_2019}.
We note that the quenched models are underdense relative to experimental references.
However, constant-volume tests at experimental densities indicate that the key
structural signatures and the central trends discussed in this study are preserved (see Section~\ref{sec:limitations}).

As shown in Figure~\ref{fig:a_teox}b, the pair-correlation functions of $a$-\ce{TeO_{1.2}}, which is the experimentally reported composition~\cite{Liu_798_2024}, flatten beyond \SI{6}{\AA}, indicating the
absence of medium- and long-range orders and confirming successful amorphization.
Peaks \textbf{A}--\textbf{C} coincide with the characteristic
short-range motifs of $a$-\ce{TeO2} (distorted \ce{TeO4} polyhedra), whereas peaks \textbf{D} and \textbf{E}
align with features of $a$-\ce{Te}. In particular, the first \ce{Te-Te} pair peak at \SI{2.77}{\AA} matches the
\ce{Te-Te} bond distance reported from Fourier-transformed EXAFS of \ce{Se}-alloyed $a$-\ce{TeO$_{1.2}$}~\cite{Liu_798_2024},
while the peak at \SI{4.28}{\AA} (\textbf{E}) is consistent with longer-range \ce{Te}--\ce{Te} correlations
expected for chain-like connectivities rather than isolated dimers. Together with the broad
ECN distribution of \ce{Te} and representative local motifs of $a$-\ce{Te} and $a$-\ce{TeO2} in Figure~\ref{fig:a_teox}c,
these structural markers support the coexistence of two amorphous domains.
This is consistent with the \ce{Te}/\ce{TeO$_x$} mixed-phase picture inferred from spectroscopy~\cite{Liu_798_2024}.

To visualize and quantify this biphasic character at the atomic level, we performed a Bader-charge analysis.
As shown in Figure~\ref{fig:a_teox}d,
we distinguished neutral \ce{Te$^0$} associated with the $a$-\ce{Te}-like network from \ce{Te$^{4+}$} characteristic of the $a$-\ce{TeO2} matrix by their Bader charge.
$a$-\ce{TeO_{1.2}} contains a substantial \ce{Te^0} population (\SI{34}{\percent}$\pm1$), consistent with
the $\sim$\SI{40}{\percent} elemental \ce{Te} fraction inferred from bulk-sensitive XANES~\cite{Liu_798_2024}.
This value exceeds the \SI{9.7}{\percent} metallic signal reported by surface-sensitive XPS~\cite{Liu_798_2024},
which we attribute to under-sampling of the buried \ce{Te}-rich network.
Across independent cells, the \ce{Te$^0$} fraction varies only modestly at
each composition, yielding $73.21\pm0.96$, $53.50\pm2.50$, $34.31\pm1.25$,
and $15.65\pm0.71$\% for $x=0.4$, 0.8, 1.2, and 1.6, respectively (Table~S5).
This reproducibility supports the robustness of oxygen-deficiency-driven \ce{Te}-rich domain formation.

\subsection{Origin of holes}

\begin{figure}[!ht]
\centering\includegraphics[width=0.45\linewidth]{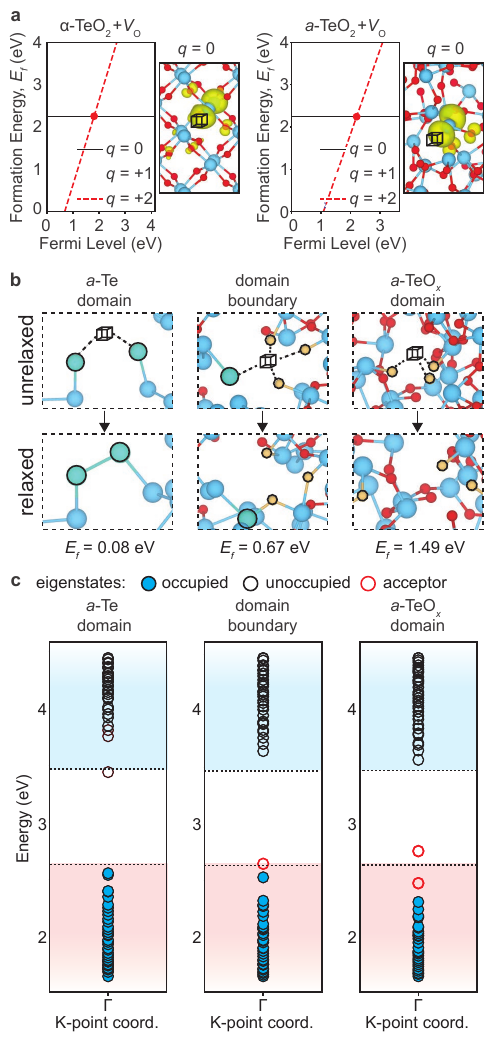}
\caption{(a) Formation energies of oxygen vacancies ($V_{\ce{O}}$) in crystalline $\alpha$-\ce{TeO2}
and $a$-\ce{TeO2} as a function of the Fermi level (referenced to the VBM), shown for charge
states $q=0$ (solid), $+1$ (dotted), and $+2$ (dashed) under \ce{O}-rich conditions; insets
depict the relaxed vacancy geometries (vacancy marked by a cube) and the charge-density
isosurface of the donor state for $q=0$ (yellow, \SI{10}{\percent} of the maximum).
(b) Representative tellurium-vacancy ($V_{\ce{Te}}$) sites in oxygen-deficient $a$-\ce{TeO$_x$}
illustrating strong environment dependence: vacancies within the $a$-\ce{Te} domain relax with
low formation energy ($E_\mathrm{f}=0.08$~eV), whereas sites at the domain boundary
($E_\mathrm{f}=0.67$~eV) or in the $a$-\ce{TeO$_x$} region ($E_\mathrm{f}=1.49$~eV) are
progressively less favorable (unrelaxed and relaxed local structures shown).
(c) Corresponding Kohn-Sham eigenvalues, highlighting that $V_{\ce{Te}}$ in oxide-like/boundary environments introduces
shallow acceptor states near the VBM, while \ce{Te} vacancies
in the $a$-\ce{Te} domain do not generate in-gap states.}
\label{fig:point_defects}
\end{figure}

We next examine which defects can supply holes in oxygen-deficient $a$-{TeO$_x$}.
We note that point defects in amorphous materials are
inherently ill-defined, as structural disorder leads to distributions
of formation energies and charge-transition levels~\cite{Strand_2306243_2023}.
Quantitative analysis therefore requires ensemble sampling over independent configurations
and inequivalent sites, while a full equilibrium concentration analysis would further
require extensive sampling of charge states and defect complexes. Here, we sample multiple
vacancy sites to capture representative variability and focus on the qualitative defect
character relevant to nonequilibrium film-growth conditions.

In oxides, oxygen vacancies ($V_{\ce{O}}$) are the most direct manifestation of oxygen deficiency and
typically act as donor defects that compensate acceptors and suppress p-type
transport, when the Fermi level is located near the VBM~\cite{Freysoldt_253_2014,Oba_060101_2018}.
Thus, we calculated $V_{\ce{O}}$ at several inequivalent bridging and non-bridging sites
spanning the dominant local environments in $a$-\ce{TeO2}.
Both formation energies of neutral $V_{\ce{O}}$ and charge-transition
levels are found to vary only modestly (within $\sim$\SI{0.15}{\eV}, available in Figure~S3 of the SI)
and closely match those in crystalline $\alpha$-\ce{TeO2}.
This indicates that $V_{\ce{O}}$ energetics in $a$-\ce{TeO2}
are governed primarily by local \ce{Te-O} bonding.
In both cases, oxygen vacancies introduce deep donor levels (Figure~\ref{fig:point_defects}a)
that localize electrons on adjacent \ce{Te}-5$p$ orbitals and pin the Fermi level at approximately $0.7$--\SI{1.1}{\eV} above the VBM,
consistent with prior studies~\cite{Huyen_044065_2024,CostaAmaral_1605_2024,Xiao_016103_2025}.

\FloatBarrier

We also calculated tellurium vacancies, $V_{\ce{Te}}$, as a potential source of holes in oxygen-poor $a$-\ce{TeO$_x$}.
In \ce{Te}-rich conditions, $V_{\ce{Te}}$
formation energies span a wide range ($0.08$--\SI{1.69}{\eV}), reflecting strong dependence on local chemistry:
they are lowest within the $a$-\ce{Te} domain and become increasingly unfavorable in \ce{TeO2}-like
environments where forming the vacancy requires breaking additional \ce{Te-O} bonds, as shown in Figure~\ref{fig:point_defects}b.
The low vacancy formation energy in the $a$-\ce{Te} phase ($0.08$–\SI{0.62}{\eV}) can be attributed to the looser packing and
greater flexibility of amorphous \ce{Te} chains, which can “heal” the vacancy through structural rearrangement,
yielding stable configurations without in-gap states. In contrast, shallow acceptor states emerge when $V_{\ce{Te}}$ resides
in the \ce{TeO2} matrix or at
the $a$-\ce{Te}/$a$-\ce{TeO2} boundary (Figure~\ref{fig:point_defects}c).
These results suggest that $V_{\ce{Te}}$ in the \ce{TeO2} regions serve as the main source of holes
in the oxygen-deficient amorphous films. Sustained p-type behavior
therefore requires these acceptor-like $V_{\ce{Te}}$ centers
to dominate over possible $V_{\ce{O}}$ compensation under nonequilibrium
deposition conditions. Experimental validation of this mechanism
could be achieved through variable-energy Positron Annihilation Lifetime Spectroscopy (PALS)
with lifetime component analysis and coincidence Doppler broadening for chemical sensitivity,
correlating the vacancy-related components with hole mobility~\cite{Hugenschmidt_547_2016}.

\subsection{Hole conduction mechanism}
We next examine how holes move through this structurally heterogeneous network.
The calculated average band gaps for $x=2.0$ and $x=1.2$
are \SI{3.51}{\eV} and \SI{1.25}{\eV}, respectively, close to experimental estimates of
\SI{3.76}{\eV}~\cite{Nayak_118_2003} and \SI{1.10}{\eV}~\cite{Liu_798_2024}.
The gap reduction at lower oxygen content is evident in the projected DOS of $a$-\ce{TeO$_{1.2}$}
(Figure~\ref{fig:a_teox}e), where \ce{Te}-$5p$ states associated with the \ce{Te}-rich domains
rise above the valence states. Accordingly, the upper valence band is
dominated by neutral \ce{Te} in the $a$-\ce{Te} subnetwork, and the corresponding charge density
concentrates along $a$-\ce{Te} percolating pathways (inset of Figure~\ref{fig:a_teox}e).
This supports hole transport through the embedded $a$-\ce{Te} network.
These trends are consistent with prior computational analysis~\cite{Liu_798_2024}, although the earlier
work found more spatially delocalized conduction pathways across the cell, likely due to methodological
differences.

To further probe the conduction mechanism, we estimate the hole
mobility-edge offset, ${\Delta}E_h$ (Figure~\ref{fig:band_scheme_amorphous}b),
using the inverse participation ratio (IPR; Methods), which approaches zero for
delocalized states and unity for states localized on a single atom.
We stress that ${\Delta}E_h$ is not a direct measure of mobility and does
not replace explicit transport calculations, but rather provides a physically
intuitive proxy for the energetic accessibility of extended valence states. Based on inspection of
charge densities (representative examples are provided in Figure~S\num{5}),
states with $\mathrm{IPR}<0.03$ are effectively delocalized in our models.
Thus, we adopt $\mathrm{IPR}=0.03$ as the mobility-edge criterion.
Sensitivity tests using $\mathrm{IPR}<0.02$ and $\mathrm{IPR}<0.04$ preserve the
same composition trend in ${\Delta}E_h$ (Figure~S\num{10}), supporting that our
conclusions do not rely on a single cutoff.
With this criterion, the average ${\Delta}E_h$ for $x = 1.2$ is \SI{0.75}{\eV},
implying that fully extended valence states are not readily accessible to hole carriers at room temperature.
Transport in $a$-\ce{TeO$_{1.2}$} should therefore remain disorder-limited within Mott’s mobility-edge framework~\cite{Mott_3075_1987}.
This can be viewed as percolation through localized or weakly localized tail states along
\ce{Te}-rich segments that are intermittently interrupted by less conductive bottlenecks.
This connectivity-assisted pathway can nonetheless yield mobilities well above those of typical amorphous oxides
(\SI{4.2}{\square\cm\per\volt\per\second})~\cite{Liu_798_2024}.

\subsection{The role of \ce{Se} doping on the hole mobility}

\begin{figure}[t!]
\centering\includegraphics[width=0.99\linewidth]{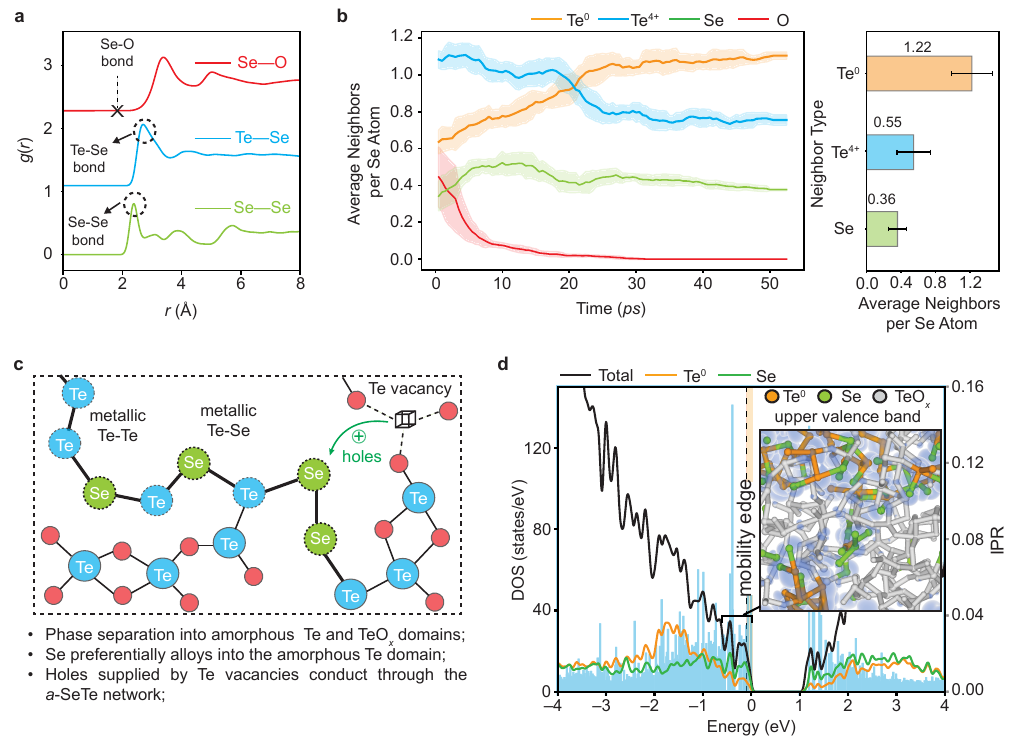}
\caption{
Atomic model and electronic properties of \ce{Se}-doped amorphous \ce{TeO$_{1.2}$} (\ce{Se$_{0.33}$TeO$_{1.2}$}).
(a) Partial pair-correlation functions, $g(r)$, for \ce{Se-O}, \ce{Te-Se}, and \ce{Se-Se} pairs, with first-neighbor peaks labeled.
(b) Statistics of \ce{Se}-centered coordination numbers along the melt–and-quench
trajectory.
Left: Averaged time evolution (10-step window); shading denotes the standard deviation within the window.
Right: final results after \SI{0}{K} HSE06 relaxation.
(c) Schematic atomic model consistent with the structural analysis,
illustrating $a$-\ce{Te}/$a$-\ce{TeO2} domain segregation and preferential
\ce{Se} alloying into the $a$-\ce{Te} conducting subnetwork.
(d) Projected DOS and energy-resolved IPR for \ce{Se$_{0.33}$TeO$_{1.2}$},
showing reduced valence-edge localization and a smaller hole mobility-edge
offset ($\Delta E_h=\SI{0.20}{\eV}$). Total DOS (black), neutral-\ce{Te$^0$} PDOS (orange),
\ce{Se} PDOS (green), and IPR (blue bars) are referenced to the VBM, with DOS and IPR
plotted on the left and right axes, respectively. Localized tail states and the mobility edge are indicated.
Inset: real-space distribution of valence-edge states
(same isosurface criteria as Figure~\ref{fig:a_teox}e), highlighting localization
on the \ce{Te$^0$} (orange) and \ce{Se} (green) framework within the oxidized matrix (grey).
}
\label{fig:se_doped_a_teox}
\end{figure}

In experiments, \ce{Se} doping significantly improves device performance in oxygen-deficient
$a$-\ce{TeO$_x$}, increasing the TFT hole mobility to \SI{15}{\square\cm\per\volt\per\second}
and the on/off ratio from $10^4$ to $10^7$~\cite{Liu_798_2024}. We modeled
the reported optimal composition, \ce{Se$_{0.33}$TeO$_{1.2}$}~\cite{Liu_798_2024}.
Since Se lies in the same column of the periodic table as Te and O,
whether Se preferentially incorporates into the $a$-\ce{Te} or $a$-\ce{TeO2} domains,
whether it behaves as an anion, a cation or a neutral atom, cannot be predicted a priori.
Our calculated partial pair-correlation functions in Figure~\ref{fig:se_doped_a_teox}a show no signature of \ce{Se-O} bonding,
whereas clear \ce{Se-Te} and \ce{Se-Se} correlations are present.
To elucidate the structure in depth, we tracked the Se-centered coordination environment
throughout the melt–quench trajectory (Figure~\ref{fig:se_doped_a_teox}b).
The initially present \ce{Se-O} coordination (red) decays rapidly during the
trajectory as oxygen atoms rebind to \ce{Te}, owing to the stronger \ce{Te-O} bonding.
This is consistent with our DFT formation energies for $\alpha$-\ce{TeO2} and \ce{SeO2} ($P4_2/mbc$),
$-298$ and \SI{-186}{\kilo\joule\per\mole}, respectively.
Instead, the number of \ce{Te$^0$} neighbors (yellow) increases, reaching 1.22 per \ce{Se} atom compared to only 0.55 \ce{Te$^{4+}$} (blue),
even though \ce{Te$^{4+}$} are more than twice as abundant in \ce{Se$_{0.33}$TeO$_{1.2}$}.
This trend indicates
that \ce{Se} preferentially incorporates into the $a$-\ce{Te} domains rather than occupying oxygen-vacancy
sites in the \ce{TeO$_x$} matrix (cf. Supplementary Figure~\num{3} of Liu et al.~\cite{Liu_798_2024}).
As illustrated in Figure~\ref{fig:se_doped_a_teox}c, this provides a concrete atomistic mechanism,
consistent with XANES/EXAFS evidence for metallic \ce{Te-Se} bonding in \ce{Se}-doped films~\cite{Liu_798_2024}.

As a result, the upper valence band of the \ce{Se}-doped sample is dominated
by $p$ states of the amorphous \ce{Te-Se} network and exhibits increased delocalization, as shown in Figure~\ref{fig:se_doped_a_teox}d.
Correspondingly, the average ${\Delta}E_h$ decreases from \SI{0.75}{\eV} to \SI{0.20}{\eV}.
When the Se doping concentration is reduced to \ce{Se$_{0.11}$TeO$_{1.2}$}, ${\Delta}E_h$ is increased to \SI{0.65}{\eV}.
This is still smaller than the undoped ${\Delta}E_h$ but substantially less than at the experimentally
optimized \ce{Se$_{0.33}$TeO$_{1.2}$} composition.
This trend aligns with recent low-temperature measurements reporting band-like transport in \ce{Se}-alloyed
$a$-\ce{TeO$_x$} TFTs~\cite{Yang_e18364_2025}.
While increasing \ce{Se} should further improve connectivity, excess \ce{Se} degrades TFT performance
via an $n$-doping effect~\cite{Liu_798_2024}, indicating an optimum rather than a monotonic mobility gain.

\subsection{Impact of oxygen composition on connectivity and mobility}\label{sec:oxygen_deficiency_knob}

\begin{figure}[t!]
\centering\includegraphics[width=0.99\linewidth]{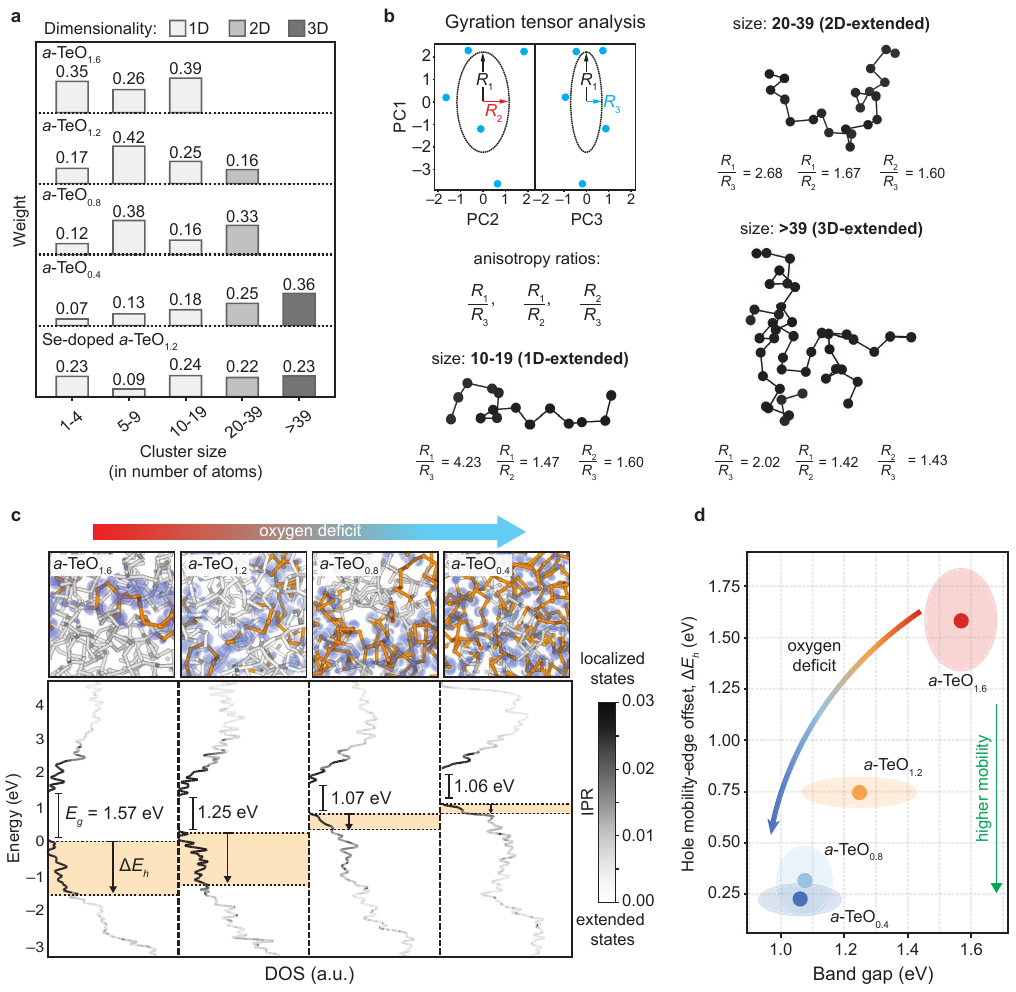}
\caption{Atomic and electronic structures in oxygen-deficient $a$-\ce{TeO$_x$}.
(a) Cluster-size distributions of the neutral conducting subnetwork (fraction of neutral atoms per size bin) for $x=1.6$, 1.2, 0.8, and 0.4,
and for \ce{Se}-doped $a$-\ce{TeO$_{1.2}$}. Bar shading indicates the dominant cluster dimensionality.
    Decreasing $x$ shifts weight toward larger clusters, indicating growth of the \ce{Te}-rich network.
(b) Gyration-tensor analysis of representative clusters,
shown in the principal-component frame with ellipsoidal radii $R_1$, $R_2$, and $R_3$.
(c) DOS evolution with oxygen deficiency, aligned by the \ce{Te^{4+}} 3$d$ core-level feature
(zero set to the VBM of $a$-\ce{TeO$_{1.6}$}). The grayscale encodes localization (IPR) and ${\Delta}E_h$
is indicated. The average band gap, $E_g$, for each composition is also shown. (d) ${\Delta}E_h$ versus band gap across compositions,
shaded ellipses indicate schematic sample-to-sample variability.}
\label{fig:a_teox_trends}
\end{figure}

Here, we analyzed the atomic connectivity through a cluster-size
decomposition and a gyration-tensor analysis of fragment shape (see Methods).
The cluster-size distributions (Figure~\ref{fig:a_teox_trends}a) shift toward larger fragments as oxygen composition $x$ decreases
from 1.6 to 0.4, indicating growth of the $a$-\ce{Te} network. \ce{Se} incorporation
at fixed $x=1.2$ produces a similar connectivity increase.
The gyration tensors, extracting principal radii $R_1$, $R_2$, and $R_3$, classify
the cluster dimensionality by anisotropy ratios (Figure~\ref{fig:a_teox_trends}b).
Small clusters (4--19 atoms) are largely 1D-extended,
consistent with \ce{Te}’s twofold coordination preference~\cite{HumeRothery_65_1930}, whereas
larger fragments evolve toward 2D motifs (20--39 atoms) and then reduced anisotropy consistent
with 3D connectivity ($>39$ atoms).
A fully connected conducting network is therefore unlikely at $x=1.2$,
supporting a disorder-limited, percolation-assisted transport picture, while stronger oxygen
deficiency promotes 3D conducting phases.
For \ce{Se}-doped $a$-\ce{TeO$_{1.2}$}, the shift toward 2D/3D-extended fragments is
consistent with the reduced mobility-edge offset and band-like signatures~\cite{Yang_e18364_2025}.
This supports the interpretation that \ce{Se} primarily enhances percolation by extending and thickening $a$-\ce{Te} conduction pathways,
analogous to oxygen deficiency.


The connectivity enhanced by oxygen deficiency is also reflected in the
DOS alignment in Figure~\ref{fig:a_teox_trends}c. As $x$ decreases, the VBM
rises monotonically, while the hole mobility-edge
offset ${\Delta}E_h$ contracts from \SI{1.59}{\eV} ($x=1.6$) to \SI{0.22}{\eV} ($x=0.4$).
Hence, these mobility gains coincide with band-gap narrowing (Figure~\ref{fig:a_teox_trends}d).
This implies that stronger oxygen deficiency can
enhance mobility by expanding the $a$-\ce{Te} network,
but also increase off-state leakage and consequently reduce TFT on/off ratios~\cite{Liu_798_2024}.
A practical workaround may be
to combine oxygen-deficiency tuning (to enhance connectivity) with controlled \ce{Se} alloying (to
adjust hole density and preserve on/off ratio).
Even then, extreme oxygen reduction is restricted by
stability: as $x\to0$ the material approaches $a$-\ce{Te}, which recrystallizes above
\SI{285}{\kelvin}~\cite{Blum_965_1974}.
Exploring such regime will require broader composition sampling.

\subsection{Generality of oxygen-deficiency-driven percolation}

\begin{figure}[!htbp]
\centering\includegraphics[width=0.6\linewidth]{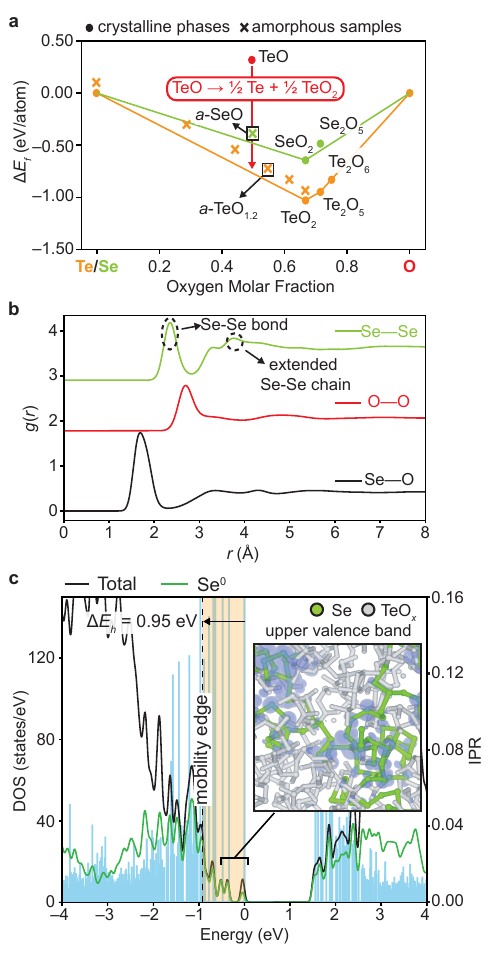}
\caption{
(a) Convex hulls of the \ce{Te}--\ce{O} (orange) and \ce{Se}--\ce{O} (green)
binaries. Circles mark crystalline phases, while crosses denote
simulated amorphous compositions (see text for details).
(b) Partial pair-correlation functions, $g(r)$, for $a$-\ce{SeO}.
(c) Total/projected DOS (neutral \ce{Se} contribution in green), on the left axis, and
energy-resolved IPR (blue bars), on the right axis, for $a$-\ce{SeO}, where the
energy zero is set to the VBM. The valence edge is dominated
by the neutral-\ce{Se} subnetwork and the mobility-edge offset $\Delta E_h$ is \SI{0.95}{\eV}.
Inset: real-space isosurface of upper-valence states (same criteria as for \ce{TeO$_x$}).
}
\label{fig:a_seox}
\end{figure}

We examine whether segregation-driven conduction could
extend to other chalcogen oxides. The \ce{Te}--\ce{O} convex hull (Figure~\ref{fig:a_seox}a)
contains no stable phases between \ce{Te} and \ce{TeO2}; for example, a hypothetical
hexagonal \ce{TeO} phase registered in the Materials Project~\cite{Jain_011002_2013}
lies far above the hull and would decompose into the end members. This suggests that
domain segregation is primarily thermodynamic in origin, as oxygen-poor compositions
favor disproportionation into elemental-like and oxide-like regions rather than
homogeneous intermediate suboxides. Microscopically, this tendency reflects
the distinct bonding preferences of oxidized and neutral tellurium: \ce{Te$^{4+}$}
favors multi-coordinated \ce{TeO2}-like environments, whereas neutral \ce{Te}
favors twofold \ce{Te-Te} chain bonding. The amorphous state should not be required
for this driving force itself, but it can kinetically trap nanoscale, interpenetrating
domains instead of allowing macroscopic crystalline phase separation. A
similar tendency is shown in the \ce{Se}--\ce{O} hull (Figure~\ref{fig:a_seox}a).
Consistent with this thermodynamic picture, our $a$-\ce{SeO} models exhibit
direct \ce{Se-Se} bonds and longer-range \ce{Se-Se} correlations indicative of chain-like
fragments (Figure~\ref{fig:a_seox}b), analogous to the $a$-\ce{Te} network
in oxygen-deficient $a$-\ce{TeO$_x$}. Their formation energies are only
modestly above the hull (96--\SI{137}{\meV\per\atom}), within
the range of experimentally accessible amorphous phases~\cite{Sun_e1600225_2016}.

Projected DOS and IPR analyses (Figure~\ref{fig:a_seox}c) show that
the upper valence band derives primarily from \ce{Se}-$4p$ states localized within the neutral \ce{Se}-rich subnetwork.
The average $\Delta E_h$, however, is $\SI{1}{\eV}$, which is larger than in $a$-\ce{TeO$_{1.2}$},
primarily due to the more localized \ce{Se}-$4p$ orbitals~\cite{Beyer_153_1971}.
This larger offset indicates that transport in $a$-\ce{SeO$_x$} is less favorable than
in $a$-\ce{TeO$_x$}. Thus, greater oxygen deficiency would likely be
required to approach band-like conductivity in $a$-\ce{SeO$_x$}.

These results should be viewed as evidence that the segregation mechanism may extend beyond tellurium oxides
and suggest that oxygen-deficiency-driven segregation may create p-type pathways in amorphous oxides when
the system
(i) lacks stable intermediate compounds between the elemental phase and the dominant bulk oxide phase and
(ii) sustains a semiconducting amorphous elemental subnetwork near the valence edge.


\subsection{Scope and limitations of our computational approach}\label{sec:limitations}
Our conclusions rely on AIMD-generated amorphous models and therefore inherit established limitations.
Finite simulation cells and rapid melt-quench protocols can underestimate densification and amplify
inter-domain boundary volume, which may overrepresent
smaller \ce{Te}-rich fragments.
Indeed, the average densities of
our models are $4.92$ and \SI{4.56}{\gram\per\cubic\cm} for $a$-\ce{TeO2} and
$a$-\ce{TeO$_{1.2}$}, underestimating experimental values of
$5.65$~\cite{Alderman_427_2019} and \SI{5.60}{\gram\per\cubic\cm}~\cite{Liu_798_2024}
by \SI{13.0}{\percent} and \SI{18.6}{\percent}, respectively.
To estimate the effect of the density underestimation, we generated $a$-\ce{TeO2} and
$a$-\ce{TeO$_{1.2}$} models at their experimental densities (NVT) and then relaxed them with HSE06.
The resulting densities remained underestimated by \SI{7.2}{\percent} and \SI{9.2}{\percent},
although pair-correlation functions and coordination statistics remained in excellent agreement with experiment.
The volume biases can shift ${\Delta}E_h$.
However, we find that, for both models at the experimental and HSE06-relaxed volumes,
the fraction of neutral \ce{Te$^0$} remains unchanged.
The band gaps of $a$-\ce{TeO2} and $a$-\ce{TeO$_{1.2}$} vary by at most \SI{0.2}{\eV},
which is within the sample standard deviation, and ${\Delta}E_h$ for $a$-\ce{TeO$_{1.2}$} changes by only \SI{0.03}{\eV}.
These results indicate that the density discrepancy does not affect the trends emphasized in this study
(see Figure~S\num{9} of the SI for the data).

Another point is that, although we generated three samples per composition to capture statistical variability,
this number may be insufficient to fully capture the spread in quantities such as band gaps and
mobility-edge offsets. Similarly, finite cells may constrain absolute domain sizes and
percolation thresholds, so we focus on reproducible signatures of segregation and
relative connectivity trends rather than exact percolation limits.
However, because all compositions and doped/undoped amorphous models
are subject to the same methodological constraints, we expect the oxygen-deficiency-driven segregation
and \ce{Se}-enhanced connectivity to remain robust.
In addition, smaller supercells are generally more prone to crystallization than to amorphization,
therefore, longer-time/larger-cell simulations should further strengthen segregation and shift
valence-edge states toward greater delocalization. The present results should thus be
viewed as a conservative baseline for the emergence of percolating \ce{Te}-rich pathways.

Although the mechanism we propose is consistent with currently available experimental results,
it requires further testing and refinement through studies involving oxygen-deficient sample preparation,
as well as spectroscopic and microscopic investigations.

\section{Conclusion}

We investigated oxygen-deficient amorphous tellurium oxides
($a$-\ce{TeO$_x$}, $x=0.0$--$2.0$), including \ce{Se}-doped $a$-\ce{TeO$_{1.2}$} using
AIMD and hybrid-functional defect calculations.
Strong oxygen depletion drives spontaneous nanoscale segregation into
interpenetrating $a$-\ce{Te} and $a$-\ce{TeO2} domains.
Notably, the two domains play complementary roles in hole conductivity;
$a$-\ce{TeO2} domains supply hole carriers through \ce{Te} vacancies,
while $a$-\ce{Te} domains form percolating \ce{Te}-$5p$ transport pathways.
We further find that \ce{Se} alloying preferentially incorporates into the \ce{Te}-rich subnetwork,
consistent with XANES and EXAFS signatures, and substantially enhances the accessibility of extended valence states.
Decreasing oxygen content offers an independent route to expanding the $a$-\ce{Te} network and lowering $\Delta E_h$,
albeit with concomitant band-gap narrowing that may increase TFT leakage.
Moreover, oxygen-deficient amorphous \ce{SeO$_x$} exhibits analogous segregation-driven valence pathways,
suggesting that this mechanism may extend beyond tellurium oxides and potentially
represent a broader route to oxygen-deficiency-driven percolation in amorphous oxides.


\section{Experimental Section}

\textbf{DFT Calculations} First-principles calculations were performed within the
framework of density functional theory (DFT) using the projector augmented
wave~\cite{Blochl_17953_1994,Kresse_1758_1999} (PAW) method
as implemented in the VASP code~\cite{Kresse_13115_1993}.
A cutoff energy of \SI{520}{\eV} was used for
variable-volume calculations to reduce Pulay stress,
while \SI{400}{\eV} was applied for constant-cell
and static calculations. Lower cutoffs were sufficient for
amorphous \ce{Te} models: \SI{227}{\eV}
during volume relaxation and \SI{175}{\eV}
for subsequent steps. The ${\Gamma}$-point was sampled to
integrate the Brillouin zone in all the calculations.
To generate the amorphous models, AIMD simulations
based on the meta-GGA r$^2$SCAN functional~\cite{Furness_8208_2020}
were done following a melt-quench protocol, summarized in
Figure~\ref{fig:a_teox}a. The AIMD trajectories were
accelerated via the on-the-fly
machine-learned force field (MLFF) framework
available in VASP\cite{Jinnouchi_014105_2019,
Jinnouchi_225701_2019,Jinnouchi_234102_2020}.
Refer to the Supporting Note~\num{2}
for information on the PAW potentials and on-the-fly training details and parameters.
The amorphous structures were modeled in cubic supercells comprising
$300$ atoms, initialized at a density of \SI{5.6}{\gram\per\cubic\cm}, which is
close to experimental values reported for $a$-\ce{TeO}$_{1.2}$~\cite{Liu_798_2024}
and $a$-\ce{TeO2}~\cite{Gulenko_14150_2014}, and used uniformly as a starting density
for all compositions (stoichiometric, oxygen-deficient, and Se-alloyed).
First, randomly-generated structures were pre-equilibrated under constant volume
(NVT ensemble) at \SI{2000}{\kelvin} for \SI{2.5}{\pico\second}
(time step \SI{5}{\femto\second}), using the Langevin thermostat~\cite{Hoover_1818_1982}.
A two-stage quenching (time step \SI{1}{\femto\second})
sequence followed: the systems were first quenched
from \SI{1500}{\kelvin} to \SI{1200}{\kelvin} in the NVT ensemble
over \SI{10}{\pico\second}, then cooled from \SI{1200}{\kelvin}
to \SI{0}{\kelvin} in the NPT ensemble, allowing the cell volume (and hence
the density) to relax self-consistently for each composition,
over \SI{40}{\pico\second}, corresponding to a cooling rate of
\SI{30}{\kelvin\per\pico\second}.
The initial high-temperature NVT segment promotes diffusion and mixing of the random structures
while avoiding unphysical volume fluctuations near the boiling point of elemental \ce{Te},
particularly in \ce{Te}-rich cells. Below \SI{1200}{\kelvin}, NPT cooling allows
controlled densification as the amorphous network condenses. To prevent unphysical cell deformation at high
temperatures, the supercells were constrained to maintain a cubic shape.
The final structures, including cell volumes, were then relaxed at \SI{0}{\kelvin}
using the r$^2$SCAN functional, followed by an additional full
variable-cell relaxation with the HSE06~\cite{Heyd_8207_2003,Krukau_224106_2006}
hybrid functional to ensure accurate electronic structure
characterization. For all geometry optimizations, energy
and force convergence criteria were set to \SI{1e-5}{\eV}
and \SI{0.03}{\eV\per\angstrom}, respectively.

\textbf{Point-defect Calculations} For the $\alpha$-\ce{TeO2} point-defect calculations, we employed a $216$-atom $3{\times}3{\times}2$ supercell
and the same convergence criteria as in the amorphous simulations. Defect formation energies, $E_f$, were computed
following established protocols~\cite{Freysoldt_253_2014,Kumagai_195205_2014}, defined for a defect $D$ in charge
state $q$ as
\begin{equation}
E_{f} = \{E[D^{q}]+E_{\text{corr}}[D^{q}]\} - E_{\text{P}} - \sum{n_{i}{\mu}_{i}} + q({\epsilon}_{\text{VBM}} + {\Delta}{\epsilon}_{\text{F}})~,
\label{eq:dfe}
\end{equation}
where $E[D^{q}]$ and $E_{\text{P}}$ are the total energies of the defective and pristine supercells, respectively; $n_i$ is the number of
atoms of element $i$ added ($n_{i} > 0$) or removed ($n_{i} < 0$); ${\mu}_{i}$ is the chemical potential of element $i$;
${\epsilon}_{\text{VBM}}$ is the VBM energy; and ${\Delta}{\epsilon}_{\text{F}}$ is the Fermi level
relative to the VBM. The correction term $E_{\text{corr}}[D^{q}]$ accounts for finite-size effects under
periodic boundary conditions and was evaluated using the extended
FNV method~\cite{PhysRevLett.102.016402,Kumagai_195205_2014,Kumagai_125202_2014}.
To compute the dielectric profiles required for these corrections, the electronic contribution was
obtained with HSE06, while the ionic part was calculated with PBEsol; for amorphous tellurium oxide,
a $100$-atom supercell was used for the ionic contribution due to computational costs.
Defect modeling and parsing were performed with PYDEFECT~\cite{pydefect,Kumagai_123803_2021}.
Transition levels were defined as the Fermi-level positions at which the most stable charge state changes.
Following defect convention, a defect is considered shallow if its transition level lies near a band
edge, facilitating ionization, and deep if it sits farther inside the band gap. For donor defects, such as \ce{O} vacancies, we specifically examined transitions from neutral to positive charge states and their
distance to the conduction band minimum. Vacancy-associated acceptor states are
identified by comparing the eigenvalues of the models with and without an oxygen vacancy,
complemented by band-decomposed charge density to assess charge localization.
The atomic structures and isosurfaces were
visualized using VESTA~\cite{Momma_1272_2011}.

\textbf{Bader Charges} The effective atomic
charges were estimated as the difference between
the number of valence electrons, $Z_{\text{val}}$, and the
Bader charge, $Q_{\text{Bader}}$, namely,
$Q_{\text{eff}} = Z_{\text{val}} - Q_{\text{Bader}}$. The Bader charges
were computed by integrating the valence pseudocharge density
within atomic compartments defined from the all-electron
charge density, using the algorithm developed
by Tang et al.~\cite{Tang_084204_2009}. To identify atoms
in the neutral state belonging to the $a$-\ce{Te} domain, we classified \ce{Te} atoms
with effective Bader charges between $-0.3$ and $+0.3$\SI{}{\elementarycharge}.
Such small fluctuations are expected from the Bader partitioning of atomic volumes and
do not compromise the distinction between reduced and oxidized \ce{Te} species.

\textbf{Inverse Participation Ratio} The inverse participation
ratio (IPR) for each Kohn--Sham eigenstate was
estimated using the projection of the atomic orbitals. Thus, the
total weight of the atom $i$ on the state $n$, $w_{i,n}$, was
calculated by summing the squared modulus of projections
over all angular momentum channels $(\ell, m)$:
\begin{equation}
    w_{i,n} = \sum_{\ell,m} \left| \langle \phi_{i,\ell m} | \psi_n \rangle \right|^2.
    \label{eq:weights}
\end{equation}
The normalized IPR is then defined as:
\begin{equation}
    \text{IPR}_n = \frac{\sum_i \left( \frac{w_{i,n}}{\sum_j w_{j,n}} \right)^2 - \frac{1}{N}}{1 - \frac{1}{N}},
    \label{eq:ipr}
\end{equation}
where $N$ is the number of atoms in the system. This normalization
ensures that $\text{IPR}_n = 0$ for a fully delocalized state
and $\text{IPR}_n = 1$ for a state entirely localized on a single atom.
The mobility edge was determined independently for each amorphous structure using the same normalized-IPR criterion; energy alignment was used only for visual comparison of DOS curves. We adopt $\mathrm{IPR}<0.03$ as the primary criterion and report a threshold-sensitivity analysis using 0.02 and 0.04 in Figure~S\num{10}.

\textbf{Gyration Tensor Analysis} To evaluate the connectivity of the hole-conducting network, we identified
connected \ce{Te}-rich fragments composed of neutral \ce{Te} atoms (and neutral \ce{Se} in doped systems).
Atoms were considered bonded when the \ce{Te-Te} or \ce{Te-Se} distance was below \SI{3.2}{\AA} or \SI{2.9}{\AA},
respectively, corresponding to the first minima in their partial pair correlation functions. For each
composition, we obtained the cluster-size distribution and weight (the relative fraction of
each size class within the \ce{Te}-rich domain) and performed shape analysis on the connected fragments.

The spatial extent and shape of each cluster were characterized by the gyration tensor. For a cluster
containing $N$ atoms with Cartesian coordinates ($\mathbf{r}^{(k)}$), the center of mass is
\begin{equation}
\mathbf{r}^{(\mathrm{CM})}
= \frac{1}{N} \sum_{k=1}^{N} \mathbf{r}^{(k)}.
\end{equation}
The components of the gyration tensor $S_{ij}$ were then computed as
\begin{equation}
S_{ij}
= \frac{1}{N}
  \sum_{k=1}^{N}
  \bigl( r_i^{(k)} - r_i^{(\mathrm{CM})} \bigr)
  \bigl( r_j^{(k)} - r_j^{(\mathrm{CM})} \bigr),
\qquad
S =
\begin{pmatrix}
S_{xx} & S_{xy} & S_{xz} \\
S_{yx} & S_{yy} & S_{yz} \\
S_{zx} & S_{zy} & S_{zz}
\end{pmatrix}.
\end{equation}
The gyration tensor $S$ is, then, diagonalized yielding three real,
non-negative eigenvalues $\lambda_1 \ge \lambda_2 \ge \lambda_3$ and the
corresponding orthonormal eigenvectors that define the principal axes of the fragment.
The associated principal radii of gyration are then obtained as
$R_i = \sqrt{\lambda_i}$ $(i = 1,2,3)$.
Thus, $R_1$, $R_2$, and $R_3$ measure the typical size of the cluster along its
longest, intermediate, and shortest dimensions, respectively, and their ratios
classify cluster geometry. We define 1D-extended (chain-like) fragments when one axis clearly dominates
($R_1/R_3 > 3$ and $R_1/R_2 > 1.6$), 2D-extended (sheet-like) when two axes are similar and
larger than the third ($1.3 < R_1/R_2 \le 1.6$ and $2 < R_1/R_3 \le 3$),
and 3D-extended (volumetric) when all three radii are comparable
($R_1/R_3 \le 2.2$ and $R_1/R_2 \le 1.4$). These thresholds, though not mathematically unique,
are physically motivated and robustly distinguish chain-like, sheet-like, and volumetric
fragments across all compositions, consistent with visual inspection of representative configurations.

\begin{suppinfo}
The Supporting Information is available free of charge at https://pubs.acs.org/doi/xxx.
It provides additional data and methodological detail that support the main text.
Supporting Note~\num{1} benchmarks exchange-correlation functionals against crystalline
tellurium and tellurium oxides to justify
r$^{2}$SCAN and HSE06. Notes \num{2}-\num{4} detail structural metrics (pair correlations, ECN analysis)
and the ML-accelerated AIMD setup.
Supporting Note~\num{5} compiles extended thermodynamic, structural, and electronic results for all $a$-\ce{TeO$_x$} and Se-alloyed
compositions (\textit{e.g.}, full DOS/IPR sets, stability trends, and pair correlation functions).
\end{suppinfo}

\section{Competing interests}
The authors declare no competing financial interests.

\section{Acknowledgments}
R.C.A. acknowledges the insightful comments from S. Jang (Tohoku University) on
machine-learning force fields and molecular dynamics.
This study was supported by the JSPS KAKENHI Grant Number 22H01755 and 25K01486,
JST FOREST Program (JPMJFR235S), and the E-IMR project at IMR, Tohoku University.
Part of the calculations were conducted using the facilities of the Supercomputer Center,
the Institute for Solid State Physics, the University of Tokyo (Project IDs: 2025-Ca-0135 and 2026-Ca-0145).

\providecommand{\latin}[1]{#1}
\makeatletter
\providecommand{\doi}
  {\begingroup\let\do\@makeother\dospecials
  \catcode`\{=1 \catcode`\}=2 \doi@aux}
\providecommand{\doi@aux}[1]{\endgroup\texttt{#1}}
\makeatother
\providecommand*\mcitethebibliography{\thebibliography}
\csname @ifundefined\endcsname{endmcitethebibliography}
  {\let\endmcitethebibliography\endthebibliography}{}

\FloatBarrier

\beginsupplement

\pagebreak
\begin{center}
{\bfseries\LARGE Supporting Information: \\ Oxygen-deficiency-driven phase segregation
enables enhanced hole transport in amorphous tellurium oxides}
\end{center}
\vspace{6pt} 

\section*{Supporting Note 1: Benchmark of exchange-correlation functionals}

Owing to the known sensitivity of \ce{Te}-based glasses to the
choice of exchange–correlation functional,\cite{Akola_134103_2012,Raghvender_174201_2022}
we benchmarked GGA, meta-GGA, and hybrid functionals against the structural and
electronic properties of $c$-\ce{Te}, $\alpha$-, $\beta$-, and $\gamma$-\ce{TeO2},
and \ce{Te2O5}, as summarized in Table~\ref{tab:xc_benchmark}. While these crystalline
phases differ from the amorphous state, they offer a computationally
affordable basis for assessing functional performance across relevant local bonding environments.
Except for trigonal \ce{Te}, best described by PBE, the meta-GGA r$^2$SCAN
functional achieved the best balance of accuracy and cost and was therefore
adopted for quenching simulations. The HSE06 hybrid
functional~\cite{Heyd_8207_2003,Krukau_224106_2006} was used
post-quench to relax the volume and atomic positions, while refining the electronic structure.

\begin{table}[H]
\centering\small
\caption{Calculated lattice parameters ($a,b,c$ in \AA), cell volume $V$ (\AA$^{3}$)
and band gap, $E_{g}$ (eV), for trigonal \ce{Te} ($c$-\ce{Te}) and
\ce{Te}–\ce{O} compounds.  Percent errors with respect to experiment are given in parentheses.}
\begin{tabular}{llccccc}
\toprule
Crystal & Functional &
$a$ (\AA) & $b$ (\AA) & $c$ (\AA) &
$V$ (\AA$^{3}$) & $E_{g}$ (eV) \\
\midrule
\multicolumn{7}{l}{\textbf{$c$-Te}}\\
& PBE      & 4.51 (1.16) & 4.51 (1.16) & 5.96 (0.70) & 104.92 (3.05) & 0.17 ($-$48.12) \\
& PBEsol   & 4.31 ($-$3.28) & 4.31 ($-$3.28) & 5.96 (0.73) & 95.94 ($-$5.77) & 0 (metal) \\
& r$^2$SCAN& 4.60 (3.19) & 4.60 (3.19) & 5.92 ($-$0.05) & 108.36 (6.43) & 0.47 (47.50) \\
& HSE06    & 4.62 (3.64) & 4.62 (3.64) & 5.80 ($-$2.03) & 107.14 (5.23) & 0.89 (177.19) \\
& Exp.~\cite{Adenis_941_1989}     & 4.46  & 4.46  & 5.92  & 101.81  & 0.32~\cite{Loferski_707_1954} \\
\midrule
\multicolumn{7}{l}{\textbf{$\alpha$-TeO\textsubscript{2}}}\\
& PBE      & 4.98 (3.43) & 4.98 (3.43) & 7.62 (0.12) & 188.84 (7.10) & 2.80 ($-$20.03) \\
& PBEsol   & 4.82 (0.21) & 4.82 (0.21) & 7.44 ($-$2.26) & 173.06 ($-$1.85) & 2.83 ($-$19.14) \\
& r$^2$SCAN& 4.84 (0.47) & 4.84 (0.47) & 7.67 (0.70) & 179.25 (1.65) & 3.41 ($-$2.60) \\
& HSE06    & 4.87 (1.12) & 4.87 (1.12) & 7.65 (0.42) & 181.04 (2.67) & 4.15 (18.43) \\
& Exp.~\cite{Lindqvist_977_1968}     & 4.81   & 4.81   & 7.62   & 176.33  & 3.50~\cite{Jain_701_1981} \\
\midrule
\multicolumn{7}{l}{\textbf{$\beta$-TeO\textsubscript{2}}}\\
& PBE      & 5.59 (2.30) & 5.75 (2.58) & 12.33 (2.43) & 396.39 (7.51) & 2.25 ($-$39.24) \\
& PBEsol   & 5.33 ($-$2.49) & 5.70 (1.66) & 11.85 ($-$1.52) & 360.02 ($-$2.36) & 2.02 ($-$45.43) \\
& r$^2$SCAN& 5.44 ($-$0.37) & 5.63 (0.32) & 12.33 (2.46) & 377.57 (2.40) & 2.76 ($-$25.49) \\
& HSE06    & 5.52 (0.95) & 5.62 (0.17) & 12.49 (3.81) & 387.03 (4.97) & 3.64 ($-$1.68) \\
& Exp.~\cite{Beyer_228_1967}     & 5.46   & 5.61   & 12.04   & 368.71   & 3.70~\cite{Shi_2006230_2021} \\
\midrule
\multicolumn{7}{l}{\textbf{$\gamma$-TeO\textsubscript{2}}}\\
& PBE      & 4.48 (2.99) & 5.11 (4.35) & 8.81 (2.75) & 201.82 (10.42) & 3.05 ($-$10.50) \\
& PBEsol   & 4.27 ($-$1.98) & 4.93 (0.67) & 8.53 ($-$0.49) & 179.46 ($-$1.81) & 3.08 ($-$9.77) \\
& r$^2$SCAN& 4.41 (1.41) & 4.87 ($-$0.51) & 8.75 (1.99) & 188.07 (2.90) & 3.63 (6.36) \\
& HSE06    & 4.47 (2.63) & 4.93 (0.59) & 8.82 (2.82) & 193.98 (6.14) & 4.37 (28.09) \\
& Exp.~\cite{ChamparnaudMesjard_1499_2000}     & 4.35   & 4.90   & 8.58   & 182.77   & 3.41~\cite{Dewan_237_2007} \\
\midrule
\multicolumn{7}{l}{\textbf{Te\textsubscript{2}O\textsubscript{5}}}\\
& PBE      & 5.51 (2.71) & 4.80 (2.21) & 8.16 (2.61) & 210.85 (8.77) & 1.68 \\
& PBEsol   & 5.46 (1.77) & 4.71 (0.38) & 8.00 (0.53) & 200.49 (3.42) & 1.82 \\
& r$^2$SCAN& 5.40 (0.63) & 4.72 (0.53) & 8.00 (0.53) & 197.79 (2.02) & 2.39 \\
& HSE06    & 5.39 (0.45) & 4.74 (0.92) & 8.06 (1.27) & 200.66 (3.50) & 3.05 \\
& Exp.~\cite{Lindqvist_643_1973}     & 5.37   & 4.70   & 7.96   & 193.86   & -- \\
\bottomrule
\end{tabular}
\label{tab:xc_benchmark}
\end{table}

\section*{Supporting Note 2: Computational Details and Machine Learning Parameters}

The projector augmented wave (PAW) potentials used in this study
are listed in Table~\ref{tab:paw_data}. We initially attempted to
run the molecular dynamics entirely using machine-learning force fields (MLFFs).
However, this led to unphysical dynamics, structural collapse, or divergent cell
volumes, likely due to the need for a much larger sampling space and longer
training times, given the disordered nature of the structures.
Additionally, the presence of multiple compositions introduced
challenges in transferring the MLFFs across different \ce{O}:\ce{Te} ratios.
We therefore switched to on-the-fly learning, triggering \textit{ab initio}
evaluations whenever the predicted error exceeded a threshold. Each quenching process
was divided into five independent and sequential runs of \SI{10}{\pico\second}, with no force
field pretraining or transfer between runs. In the production quench trajectories, approximately \SI{3}{\percent} of the steps required
explicit \textit{ab initio} evaluations. This adaptive sampling was important because the trajectories traverse chemically distinct
metallic \ce{Te}-like and ionic \ce{TeO2}-like environments involving bond breaking and formation. While a higher number of
\textit{ab initio} evaluations was performed, this approach yielded stable trajectories
across all compositions without relying on transferable MLFFs, ensuring simulations
consistent with experiments, as discussed in Supporting Note~\num{3}.
The VASP machine-learning parameters
are summarized in Table~\ref{tab:ml_parameters}.

\begin{table}[H]
\caption{Information on the PAW data sets adopted in this study.}
\label{tab:paw_data}
\centering
\begin{tabular}{ccc}
\hline
\hline
Element & VASP symbol & Valence orbitals \\ \bottomrule
O & O & (2s)$^2$ (2p)$^4$  \\
Te & Te & (5p)$^4$ (5s)$^2$  \\
Se & Se & (4s)$^2$ (4p)$^4$ \\ \bottomrule
\end{tabular}
\end{table}

\begin{table}[H]
\centering\small
\caption{Machine-learning INCAR parameters used for
on-the-fly accelerated \textit{ab initio} molecular dynamics.}
\label{tab:ml_parameters}
\begin{tabular}{lc}
\toprule
Flag &  Value used \\
\midrule
\texttt{ML\_LMLFF}          & \texttt{.TRUE.} \\
\texttt{ML\_MODE}           & train \\
\texttt{ML\_ISTART}           & 0 \\
\texttt{ML\_RCUT1}           & 8.0 \\
\texttt{ML\_SION1}           & 0.5 \\
\texttt{ML\_MRB1}           & 12 \\
\texttt{ML\_DESC\_TYPE}           & 0 \\
\texttt{ML\_RCUT2}           & 5.0 \\
\texttt{ML\_SION2}           & 0.5 \\
\texttt{ML\_MRB2}            & 8 \\
\texttt{ML\_LMAX2}           & 3 \\
\texttt{ML\_ICRITERIA}           & 1 \\
\texttt{ML\_IUPDATE\_CRITERIA}           & 1 \\
\texttt{ML\_CTIFOR}           & 0.002 \\
\texttt{ML\_SCLC\_CTIFOR}           & 0.6 \\
\texttt{ML\_CX}           & 0 \\ \bottomrule
\end{tabular}
\end{table}

\section*{Supporting Note 3: Assessment of structural order and local atomic environments}

To assess structural order in our amorphous models, we used
the pair-correlation (radial distribution) function,
$g_{AB}(r)$, or just $g(r)$, which quantifies the probability of finding
a particle of type $B$ at a distance $r$ from a particle of type $A$, relative to
an ideal gas with density $\rho_B$. The $g_{AB}(r)$ function
is computed as:
\begin{equation}
g_{AB}(r) = \frac{1}{4\pi r^2 \rho_B N_A \Delta r}
\sum_{i \in A} \sum_{j \in B}
\Theta\left( r + \frac{\Delta r}{2} - r_{ij} \right)
\Theta\left( r_{ij} - \left( r - \frac{\Delta r}{2} \right) \right),
\end{equation}
where $N_A$ is the number of $A$ atoms, $r_{ij}$ the interatomic distance, $\Delta r$
the histogram bin width, and $\Theta$ the Heaviside step function selecting pairs within the shell centered at $r$.
Complementarily, to describe local atomic environments in disordered systems,
we used the effective coordination number (ECN)~\cite{Hoppe_25_1970,Hirshfeld_129_1977},
which is particularly suitable for amorphous materials, where
bond lengths vary significantly. The ECN assigns a
continuous weight to each bond rather than relying on a fixed cutoff,
for an atom $i$ it is defined as:
\begin{equation}
\text{ECN}_i = \sum_j \exp\left[ 1 - \left( \frac{d_{ij}}{d^{i}_{av}} \right)^6 \right],
\end{equation}
where $d_{ij}$ is the distance between atoms $i$ and $j$, and $d^{i}_{av}$ is
the atom-specific average bond length
computed self-consistently via:\cite{DaSilva_023502_2011}
\begin{equation}
d^{i}_{\mathrm{av}} =
\frac{\sum_j d_{ij} \exp\left[ 1 - \left( \frac{d_{ij}}{d^{i}_{av}} \right)^6 \right]}
{\sum_j \exp\left[ 1 - \left( \frac{d_{ij}}{d^{i}_{av}} \right)^6 \right]}.
\end{equation}
The iteration begins with the shortest bond length from atom $i$,
and converges when the change in $d^{i}_{av}$ between successive
steps is smaller than \SI{1e-4}{\AA}. This formulation allows
for a smooth distinction between strong and weak bonding
contributions, enabling consistent coordination
analysis in both ordered and disordered systems.

\section*{Supporting Note 4: Structural features of amorphous \ce{Te} and \ce{TeO2}}

\begin{figure*}[t!]
\centering\includegraphics[width=0.99\linewidth]{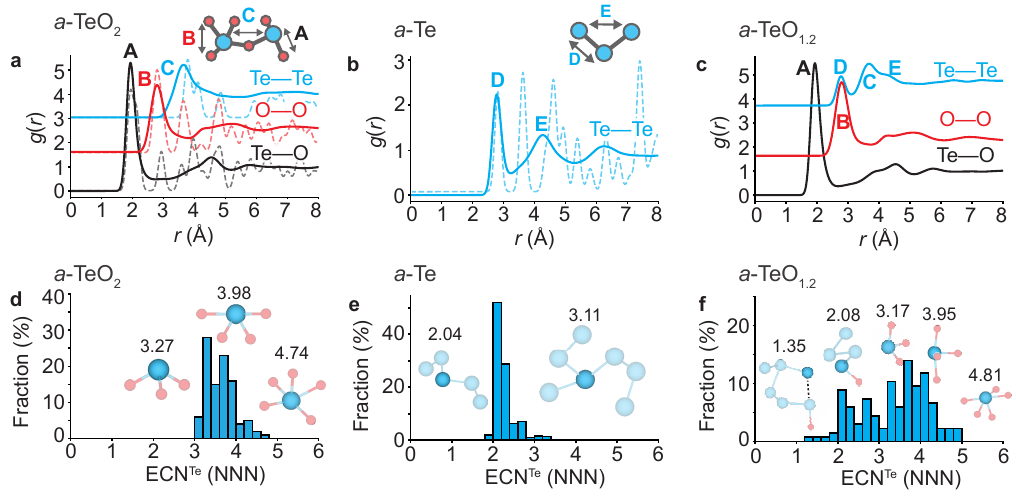}
\caption{Pair correlation functions and distributions of tellurium coordination numbers.
(a--c) Pair correlation functions, $g(r)$, of (a) $a$-\ce{TeO2},
(b) $a$-\ce{Te}, and (c) $a$-\ce{TeO_{1.2}}; Radial atomic distributions
of \ce{Te}--\ce{O}, \ce{O}--\ce{O}, and \ce{Te}--{Te} pairs for the amorphous samples are displayed in
solid lines; those for crystalline $\alpha$-\ce{TeO2} (paratellurite) and \ce{Te} (trigonal)
are shown in dashed lines. The peaks are labeled and assigned to
atomic pairs of the structural fragments shown in the
insets. The curves are shifted vertically for clarity.
(d--f) Histograms of tellurium effective
coordination number, ECN$^{\ce{Te}}$, in number of nearest neighbors (NNN).
The insets show fragments of the amorphous structures with corresponding ECN$^{\ce{Te}}$.}
\label{fig:pcf_ecn}
\end{figure*}

Figure~\ref{fig:pcf_ecn} summarizes the structural characteristics of the amorphous models,
presenting pair correlation functions with each peak assigned to
its bonding environment, and histograms of the tellurium effective
coordination numbers, ECN$^{\ce{Te}}$; the corresponding $g(r)$ peak
positions and the average tellurium ECN, $n_{\text{Te}}$, in
number of nearest neighbors (NNN) are summarized in Table~\ref{tab:gr_peaks}.
In $a$-\ce{TeO2}, peak \textbf{A} (\SI{1.94}{\AA}) corresponds
to the \ce{Te-O} bond, while peaks \textbf{B} (\SI{2.79}{\AA}) and
\textbf{C} (\SI{3.66}{\AA}) arise from intra-unit \ce{O-O} and
inter-unit \ce{Te-Te} distances, respectively. These peak positions
closely match experimental values of \SI{1.93}{\AA}, \SI{2.79}{\AA},
and \SI{3.55}{\AA} from high-resolution X-ray pair distribution
functions of \ce{TeO2} glasses~\cite{Alderman_427_2019} (Table~\ref{tab:gr_peaks}).
The ECN$^{\ce{Te}}$ distribution (Figure~\ref{fig:pcf_ecn}d) reveals
three- to five-fold coordination environments, consistent with
prior theoretical models.\cite{Gulenko_14150_2014,Raghvender_174201_2022} The computed average of
$3.63$~{NNN} falls slightly below the experimental value of 4.0,\cite{Alderman_427_2019}
but lies within the reported range of 3.6--4.0,\cite{Barney_2312_2013,Raghvender_174201_2022,Gulenko_14150_2014,Marple_183_2019}
reflecting the inherent ambiguity in defining \ce{Te-O} cutoffs in disordered networks.
In $a$-\ce{Te}, peaks \textbf{D} (\SI{2.79}{\AA}) and \textbf{E} (\SI{4.27}{\AA}) correspond
to the first and second \ce{Te-Te} neighbors within chain-like motifs, in agreement with
electron diffraction data.\cite{Ichikawa_707_1973} A minor population of threefold coordinated
\ce{Te} atoms (Figure~\ref{fig:pcf_ecn}e) yields an average coordination number of $2.26$~{NNN}, consistent with prior
\textit{ab initio} molecular dynamics estimates ($2.12$--$2.69$).\cite{Akola_134103_2012}
This strong agreement across both stoichiometric endmembers supports the fidelity of
our methodology in capturing the local environments of intermediate oxygen-deficient phases.

\begin{table}[H]
\centering
\begin{tabular}{lccccccc}
\hline
\multirow{2}{*}{} & \multirow{2}{*}{} & \multicolumn{5}{c}{$g(r)$ peaks} & \multirow{2}{*}{$n_{\text{Te}}$} \\ 
 & & A & B & C & D & E & \\
\hline
\multirow{2}{*}{$a$-\ce{TeO2}}
    & This work                     & \num{1.94} & \num{2.79} & \num{3.66} & -- & -- & \num{3.63} \\
    & Exp.~\cite{Alderman_427_2019} & \num{1.93} & \num{2.79} & \num{3.55} & -- & -- & \num{4} \\
\hline
$a$-\ce{TeO_{1.2}}
    & This work                     & \num{1.94} & \num{2.80} & \num{3.67} & \num{2.77} & \num{4.28} & \num{3.05} \\
\hline
\multirow{2}{*}{$a$-\ce{Te}}
    & This work                     & -- & -- & -- & \num{2.79} & \num{4.27} & \num{2.19} \\
    & Exp.~\cite{Ichikawa_707_1973} & -- & -- & -- & \num{2.80} & \num{4.25} & \num{1.8} \\
\hline
\end{tabular}
\caption{Pair-correlation function, $g(r)$, peak positions (in \AA) and average effective coordination number of \ce{Te} ($n_{\text{Te}}$) for the investigated structures, compared to experimental data.}
\label{tab:gr_peaks}
\end{table}

\section*{Supporting Note 5: Extra structural and electronic data}

\begin{figure*}[h!]
\centering\includegraphics[width=0.99\linewidth]{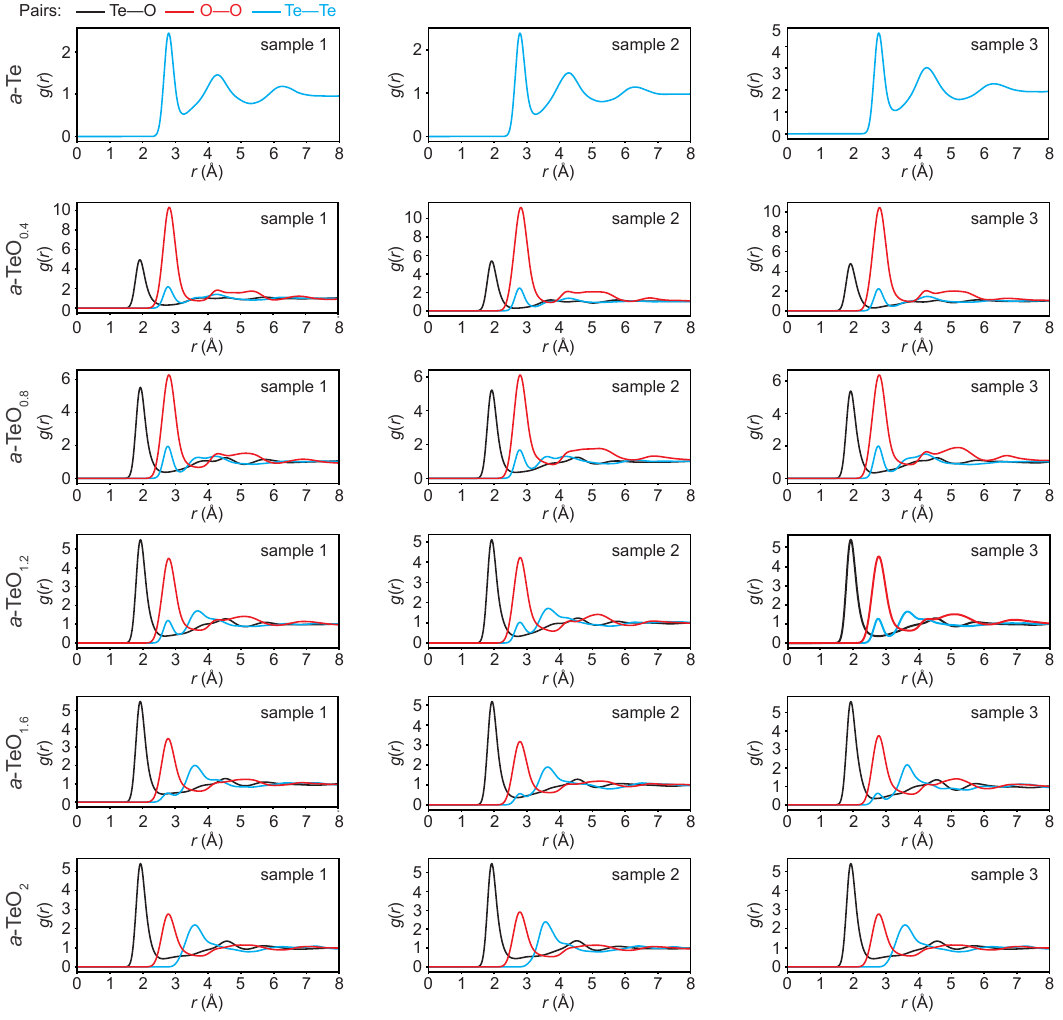}
\caption{Pair-correlation functions, $g(r)$, for \ce{Te}--\ce{O} (black),
\ce{O}--\ce{O} (red), and \ce{Te}--\ce{Te} (blue) atom pairs, computed for independent amorphous samples
at each composition from $x=0$ ($a$-\ce{Te}) to $x=2$ ($a$-\ce{TeO2}). All structures were
generated by \textit{ab initio} melt-quench simulations under identical conditions. The
systematic evolution of peak positions and intensities with oxygen content
confirms the reproducibility and structural consistency across the compositional series.}
\label{fig:pcf_SI}
\end{figure*}

\begin{figure*}[h!]
\centering
\includegraphics[width=0.9\linewidth]{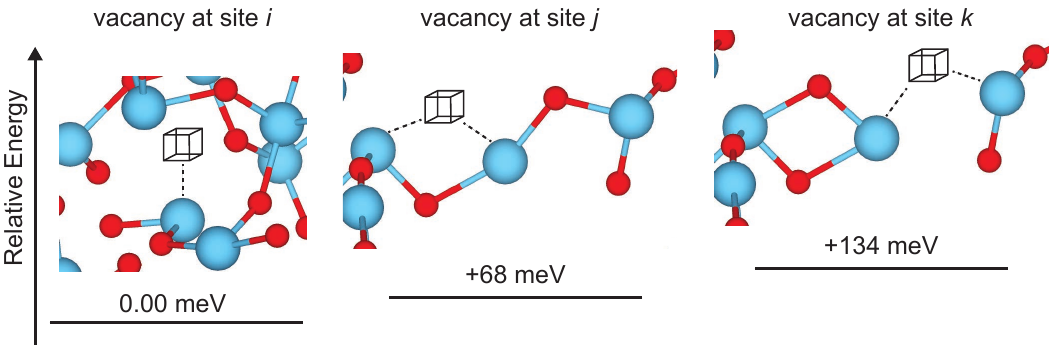}
\caption{
Local atomic environments for three distinct vacancy sites ($i$--$k$), with
relative energies of $0.00$, $+68$, and $+$\SI{134}{\meV} per vacancy,
respectively. Vacancies are indicated by transparent cubes, whereas \ce{Te} and
\ce{O} atoms are shown in blue and red.}
\label{fig:vacancy_sites}
\end{figure*}

\begin{figure*}[h!]
\centering
\includegraphics[width=0.99\linewidth]{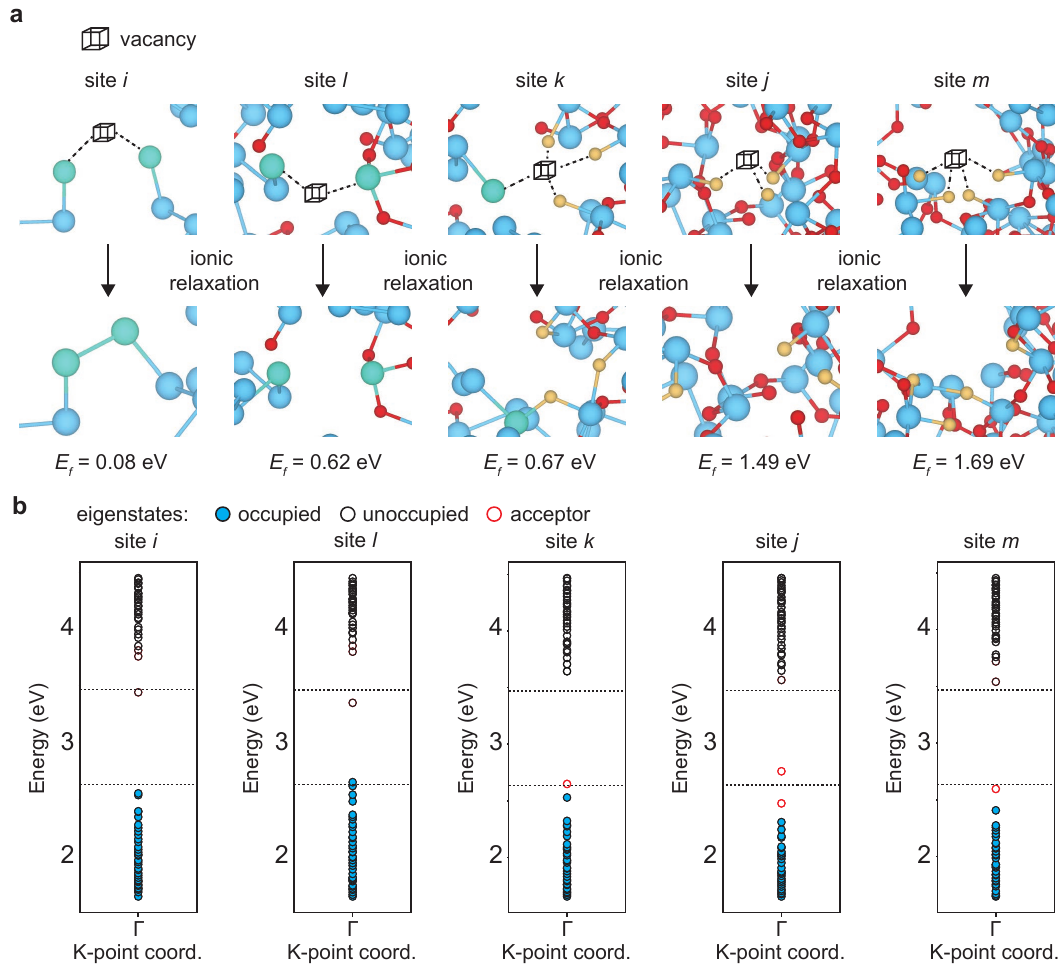}
\caption{(a) Local bonding environments of five non-equivalent \ce{Te} sites
in an $a$-\ce{TeO$_{0.8}$} model before and after ionic relaxation, highlighting
the \ce{Te}-vacancy configurations and their corresponding formation energies. Tellurium
atoms are shown in blue, oxygen in red, with \ce{Te} atoms neighboring the vacancy
highlighted in cyan and oxygen in orange. (b) Corresponding electronic eigenstates for
each site; filled and empty circles denote occupied and unoccupied states, respectively.
Red circles denote \ce{Te}-vacancy states. The dotted lines indicate the valence- and
conduction-band edges of the defect-free supercell.}
\label{fig:te_vacancies}
\end{figure*}

\begin{figure*}[h!]
\centering
\includegraphics[width=0.99\linewidth]{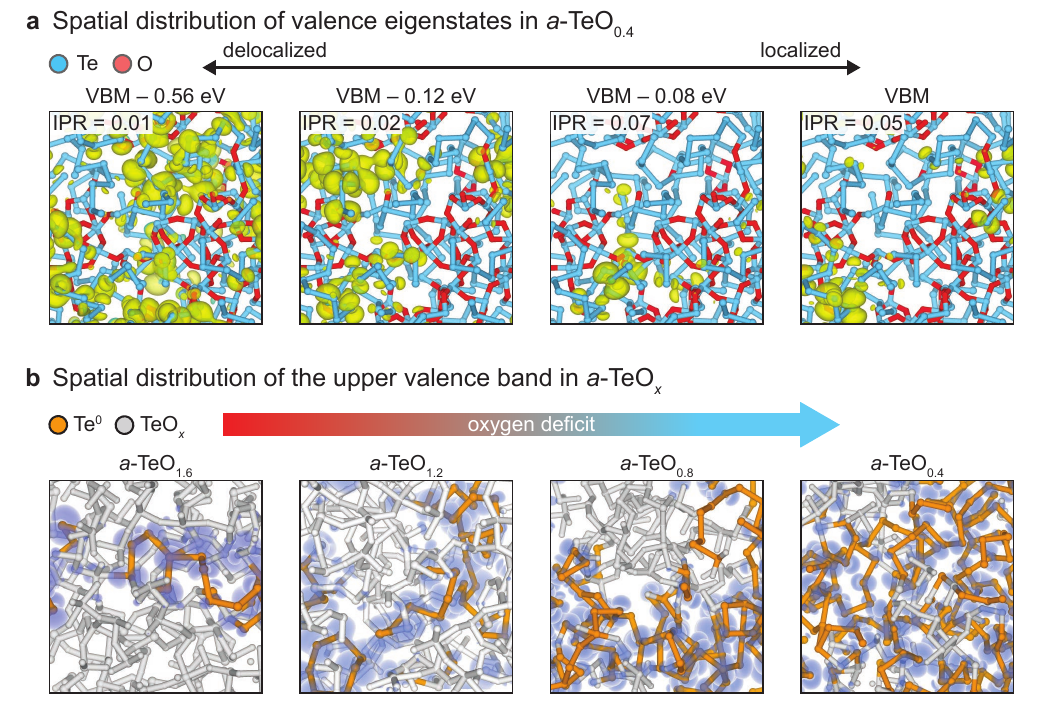}
\caption{
(a) Isosurface plots of selected valence-band eigenstates in $a$-\ce{TeO$_{0.4}$},
shown with their energy relative to VBM and corresponding IPR. States with $\text{IPR} < 0.03$, deeper in the valence
band, are typically delocalized. Yellow isosurfaces indicate the real-space distribution
of the wavefunctions; \ce{Te} and \ce{O} atoms are shown in blue and red, respectively.
(b) Real-space distribution of the upper valence-band states (in blue) across compositions
from $x = 1.6$ to $0.4$, highlighting the transition from a percolative to a
uniform network. The upper valence band, concentrated within the neutral
\ce{Te}-rich subnetwork (orange), progressively extends with decreasing oxygen content.
The isosurface corresponds to the states
within \SI{0.5}{\eV} below the valence band maximum (VBM), plotted at an isovalue of
\SI{5}{\percent} of the maximum charge density.}
\label{fig:ipr_vbm_ed}
\end{figure*}

\begin{figure*}[h!]
\centering\includegraphics[width=0.99\linewidth]{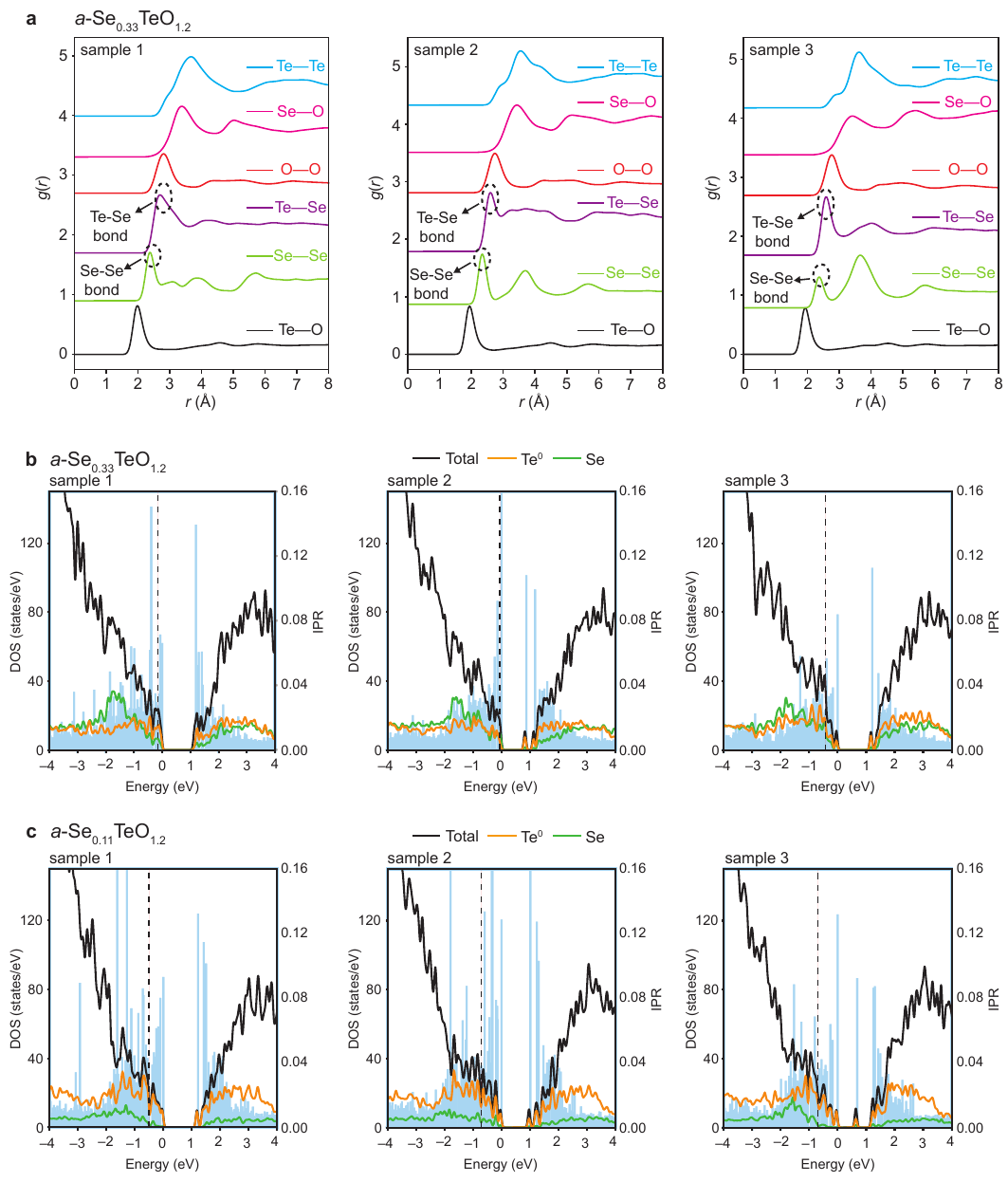}
\caption{(a) Partial pair correlation functions $g(r)$ for the \ce{Se$_{0.33}$TeO$_{1.2}$} samples.
Dashed circles highlight the main interatomic correlations.
Vertical offsets applied for clarity.
(b, c) Projected DOS and energy-resolved IPR for samples of
(b) \ce{Se$_{0.33}$TeO$_{1.2}$} and (c) \ce{Se$_{0.11}$TeO$_{1.2}$}, showing that the
upper valence band is dominated by the neutral \ce{Te$^0$} and \ce{Se} subnetwork. The vertical dashed line
indicates the estimated position of the mobility edge.}
\label{fig:si_se_doped_teox}
\end{figure*}

\begin{figure*}[h!]
\centering\includegraphics[width=0.99\linewidth]{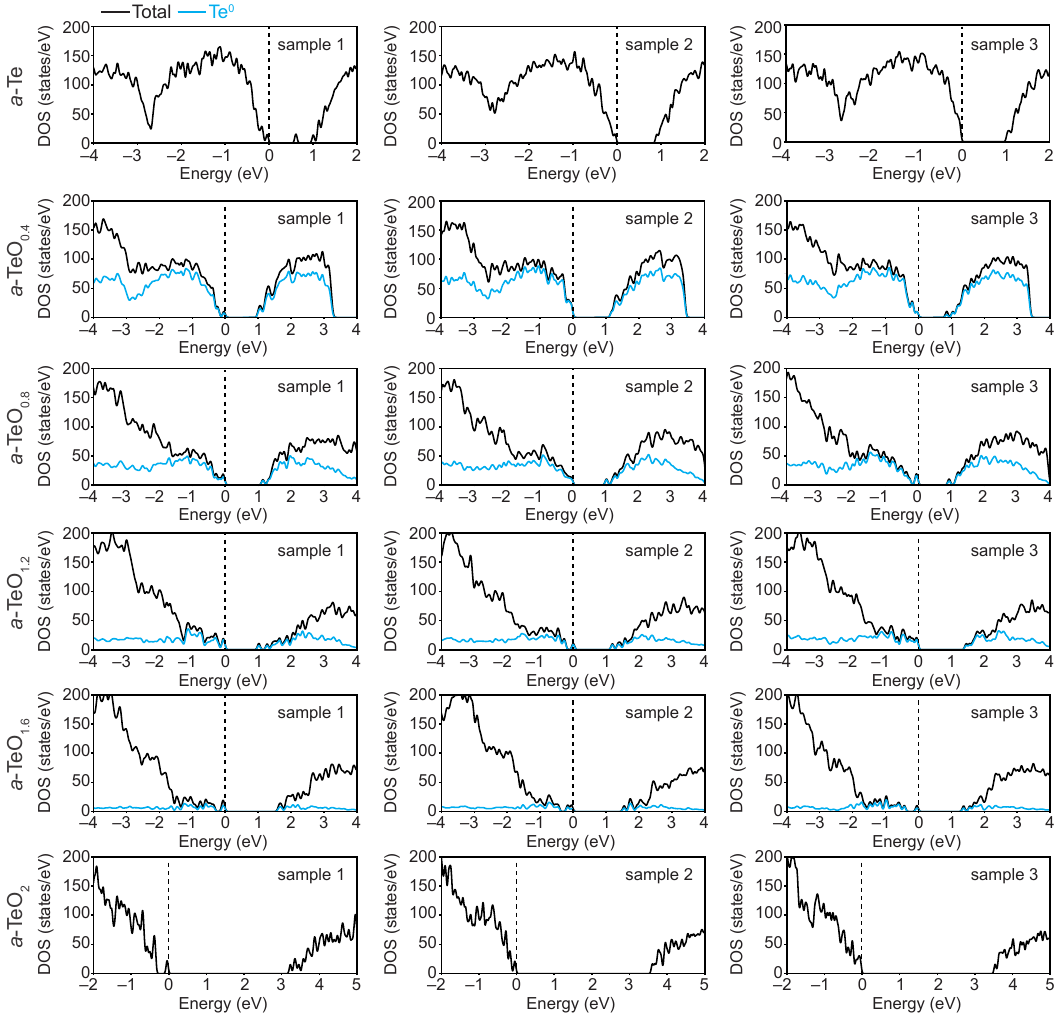}
\caption{Total (black) and projected (blue) DOS for independent amorphous samples
at each composition from $x=0$ ($a$-\ce{Te}) to $x=2$ ($a$-\ce{TeO2}). The projected DOS corresponds
to the contribution from neutral \ce{Te$^0$} atoms. The energy
is given relative to the valence band maximum (VBM), represented by
the vertical dashed line. The evolution of the valence and conduction band edges with
oxygen content reflects the transition
from semimetallic to insulating behavior across the series.}
\label{fig:dos_teox_SI}
\end{figure*}

\begin{figure*}[h!]
\centering\includegraphics[width=0.99\linewidth]{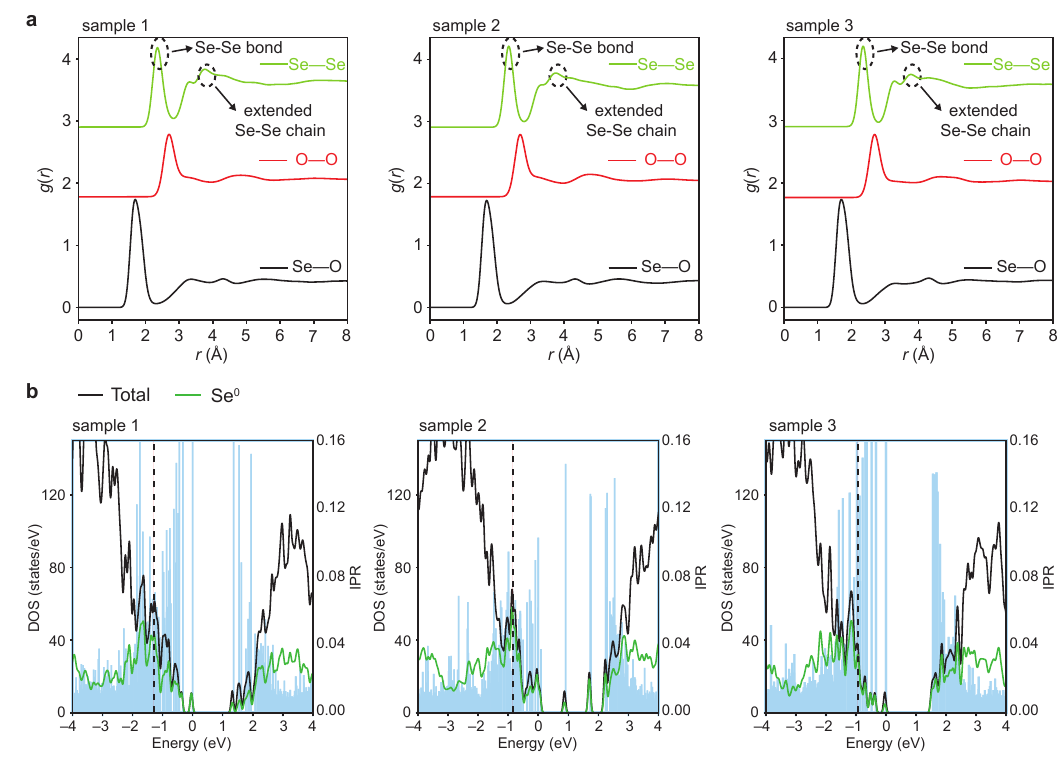}
\caption{Structural and electronic properties of oxygen-deficient amorphous \ce{SeO$_x$} models.
(a) Partial pair-correlation functions, $g(r)$, for \ce{Se-O}, \ce{O-O}, and \ce{Se-Se}
pairs in three independently generated samples. Dashed circles highlight the first-neighbor \ce{Se-Se}
bond and longer-range correlations associated with extended \ce{Se-Se} chain motifs.
(b) Total DOS (black line), neutral-\ce{Se$^0$}-projected DOS (green line), and IPR (blue bars) for the same samples.
Energies are referenced to the VBM; DOS and IPR are plotted on the left and right axes, respectively.}
\label{fig:si-a_seox}
\end{figure*}

\begin{figure*}[h!]
\centering\includegraphics[width=0.99\linewidth]{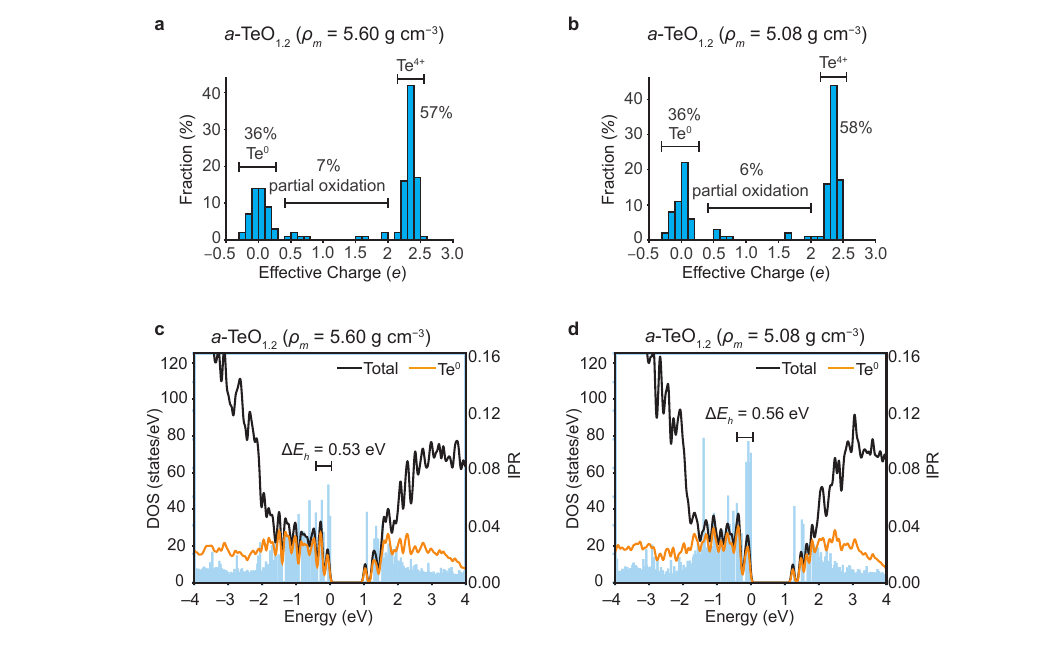}
\caption{(a,b) Distributions of \ce{Te} effective charge (from Bader analysis) for $a$-\ce{TeO$_{1.2}$} models
prepared at the experimental density ($\rho_m=\SI{5.60}{\gram\per\cubic\cm}$) and at the lower density
obtained after HSE06 relaxation ($\rho_m=\SI{5.08}{\gram\per\cubic\cm}$). The bracketed ranges indicate
the charge windows used to classify neutral \ce{Te$^0$}, partially oxidized \ce{Te$^{\delta+}$}, and fully
oxidized \ce{Te$^{4+}$} populations; the corresponding fractions are labeled. (c,d) Total DOS (black) and
\ce{Te$^0$}-projected contribution (orange) with energy-resolved IPR (blue bars) for the same models
(energies referenced to the VBM). The hole mobility-edge offset, $\Delta E_h$, extracted from the IPR criterion,
remains essentially unchanged upon densification ($0.53$ vs \SI{0.56}{\eV}), indicating that the
inferred accessibility of extended valence states is robust to the density variation considered here.}
\label{fig:a_teox_density}
\end{figure*}

\begin{figure*}[h!]
\centering\includegraphics[width=0.70\linewidth]{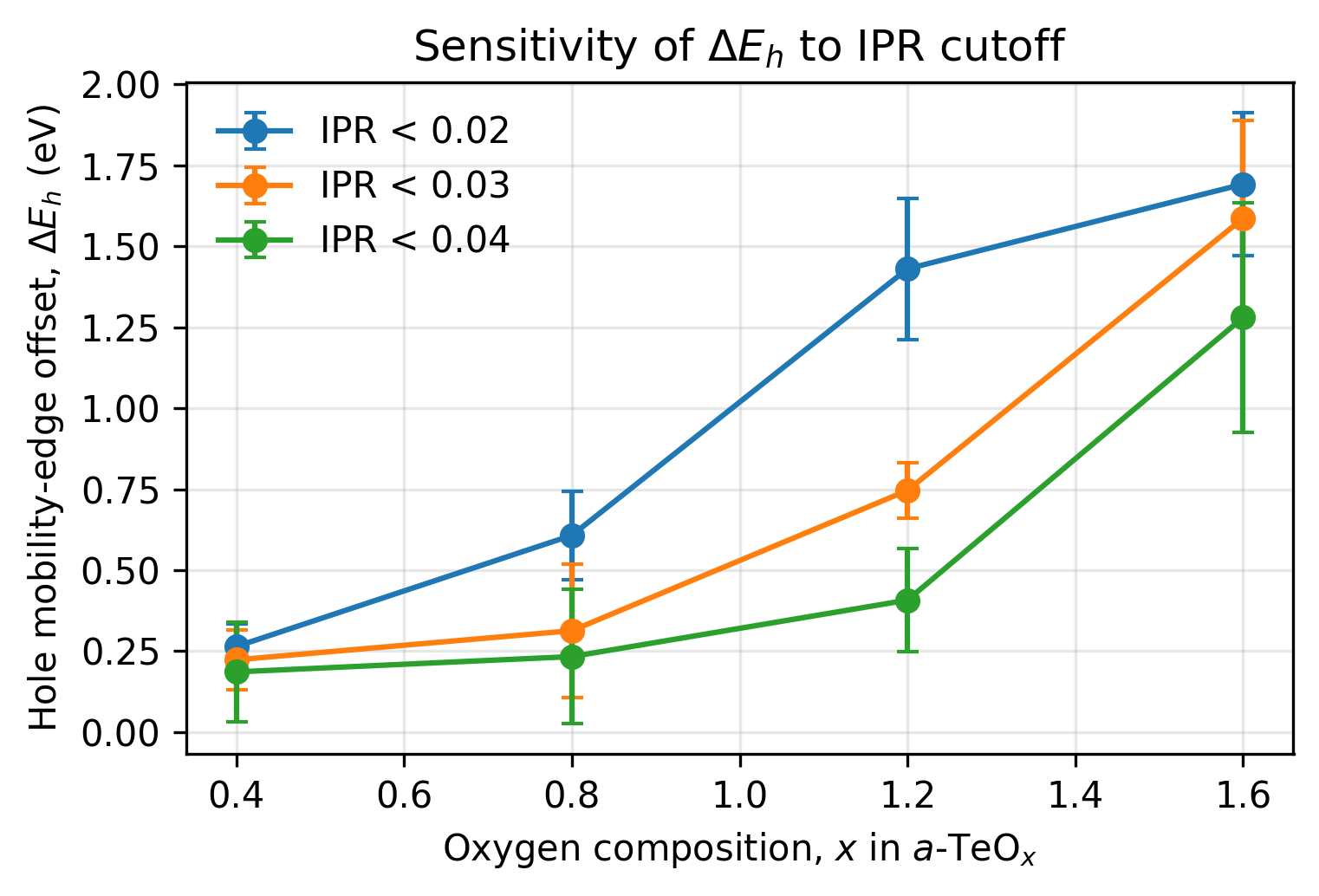}
\caption{Sensitivity of the hole mobility-edge offset, $\Delta E_h$, to the IPR threshold used to define extended states. Mean values and standard deviations are computed from three independently generated amorphous samples per composition. Although the absolute $\Delta E_h$ values depend on the chosen cutoff, the trend of decreasing $\Delta E_h$ with increasing oxygen deficiency is preserved for all criteria.}
\label{fig:ipr_sensitivity}
\end{figure*}

This section provides additional structural and electronic details of the amorphous \ce{TeO$_x$}
and \ce{Se}-alloyed samples discussed in the main text. Figure~\ref{fig:pcf_SI} shows the
pair-correlation functions, $g(r)$, for \ce{Te}–\ce{Te},
\ce{Te}–\ce{O}, and \ce{O}–\ce{O} pairs across all compositions, confirming structural
consistency and reproducibility across samples. Figure~\ref{fig:vacancy_sites} shows the
relative stability of oxygen vacancies at three distinct sites in amorphous \ce{TeO$_x$}; the
lowest-energy site was used for the charged-defect analysis. The energetics of \ce{Te} vacancies
are shown in Figure~\ref{fig:te_vacancies}, together with electronic eigenstates exhibiting
vacancy-induced empty states near the VBM, consistent with their acceptor-like character.
Figure~\ref{fig:ipr_vbm_ed} shows the spatial distribution of selected eigenstates and the
associated inverse participation ratio (IPR), together with the evolution of upper-valence-band
localization across the amorphous compositional series.
Figure~\ref{fig:si_se_doped_teox} extends the structural and electronic analysis to Se-doped
$a$-\ce{TeO$_{1.2}$} samples, reinforcing the reproducibility
and revealing differences in bonding environments. The corresponding total and projected
density of states (DOS) for all $a$-\ce{TeO$_x$} compositions are shown in
Figure~\ref{fig:dos_teox_SI}, where contributions
from neutral \ce{Te$^0$} highlight their role in shaping the upper valence band.
For the oxygen-deficient amorphous selenium oxide ($a$-\ce{SeO}), the corresponding
pair-correlation and DOS results are summarized in Figure~\ref{fig:si-a_seox}. The effective charge
histogram and DOS for $a$-\ce{TeO$_{1.2}$} having the experimental and the HSE06-relaxed density are shown in
Figure~\ref{fig:a_teox_density}, where minor differences imply little effect of the density over the electronic
properties trends established in the study. Figure~\ref{fig:ipr_sensitivity} summarizes the
sensitivity of $\Delta E_h$ to the IPR threshold used to define extended states.  Table~\ref{tab:hull_bandgap}
contains the formation energies relative to the convex hull, $E_{\text{hull}}$, mass densities,
band gaps, neutral \ce{Te$^0$} fractions, and hole mobility-edge offsets, $\Delta E_h$, for representative samples,
supporting their thermodynamic metastability and tunable transport characteristics. Finally, Table~\ref{tab:gyration_avg}
summarizes the gyration tensors properties of the cluster groups, the principal radii $R_1$, $R_2$, and $R_3$, and
dimensionality based on the anisotropy ratios.

\clearpage

\FloatBarrier
\begin{table}[H]
\centering\small
\caption{Formation energies relative to the convex hull, $E_{\text{hull}}$ (meV/atom), mass density (\SI{}{\gram\per\cubic\cm}),
band gap, $E_g$ (eV), neutral \ce{Te$^0$} fraction (\%), and hole mobility edge
offsets, $\Delta E_h$ (eV), for representative amorphous \ce{TeO$_x$} samples and
Se-alloyed $a$-\ce{TeO$_{1.2}$}}.
\begin{tabular}{lcccccc}
\toprule
    Composition     & Sample &
$E_{\text{hull}}$ (meV/atom) & density (\SI{}{\gram\per\cubic\cm}) & $E_g$ (eV) & \ce{Te$^0$} (\%) & ${\Delta}E_h$ (eV) \\
\midrule
\multicolumn{7}{l}{\textbf{$a$-\ce{TeO$_{0.4}$}}}\\
& 1       & 145.87 & 4.20 & 1.06 & 71.96 &  0.30 \\
& 2       & 134.74 & 4.41 & 1.22 & 73.36 &  0.12 \\
& 3       & 129.16 & 4.48 & 0.90 & 74.30 &  0.25 \\
& Average & 136.59 & 4.37 & 1.06 & 73.21 &  0.22 \\
\midrule
\multicolumn{7}{l}{\textbf{$a$-\ce{TeO$_{0.8}$}}}\\
& 1       & 137.33 & 4.37 & 1.19 & 52.10 &  0.41 \\
& 2       & 139.99 & 4.47 & 1.06 & 51.50 &  0.08 \\
& 3       & 121.52 & 4.45 & 0.98 & 56.89 &  0.45 \\
& Average & 132.95 & 4.43 & 1.07 & 53.50 &  0.31 \\
\midrule
\multicolumn{7}{l}{\textbf{$a$-\ce{TeO$_{1.2}$}}}\\
& 1       & 115.72 & 4.49 & 1.06 & 33.82 &  0.75 \\
& 2       & 120.24 & 4.68 & 1.19 & 33.09 &  0.66 \\
& 3       & 108.38 & 4.51 & 1.49 & 36.03 &  0.83 \\
& Average & 114.78 & 4.56 & 1.25 & 34.31 &  0.75 \\
\midrule
\multicolumn{7}{l}{\textbf{$a$-\ce{TeO$_{1.6}$}}}\\
& 1       & 114.20 & 4.66 & 1.68 & 14.78 &  1.56 \\
& 2       & 117.34 & 4.84 & 1.60 & 15.65 &  1.30 \\
& 3       & 99.96  & 4.61 & 1.42 & 16.52 &  1.90 \\
& Average & 110.50 & 4.70 & 1.57 & 15.65 &  1.59 \\\midrule
\multicolumn{7}{l}{\textbf{$a$-\ce{Se$_{0.11}$TeO$_{1.2}$}}}\\
& 1       & -- & 4.50 & 1.22 & -- &  0.50 \\
& 2       & -- & 4.68 & 1.00 & -- &  0.75 \\
& 3       & -- & 4.79 & 1.26 & -- &  0.70 \\
& Average & -- & 4.66 & 1.16 & -- &  0.65 \\
\multicolumn{7}{l}{\textbf{$a$-\ce{Se$_{0.33}$TeO$_{1.2}$}}}\\
& 1       & -- & 4.87 & 1.19 & -- &  0.15 \\
& 2       & -- & 5.02 & 0.87 & -- &  0.06 \\
& 3       & -- & 4.96 & 1.22 & -- &  0.38 \\
& Average & -- & 4.95 & 1.09 & -- &  0.20 \\\midrule
\multicolumn{7}{l}{\textbf{$a$-\ce{SeO}}}\\
& 1       & 97.20 & 3.24 & 1.36 & -- &  1.24 \\
& 2       & 92.34 & 3.20 & 1.75 & -- &  0.85 \\
& 3       & 87.11 & 3.19 & 1.56 & -- &  0.95 \\
& Average & 92.22 & 3.22 & 1.56 & -- &  1.01 \\
\bottomrule
\end{tabular}
\label{tab:hull_bandgap}
\end{table}

\begin{table}[h!]
\centering
\caption{Average anisotropy ratios, $\frac{R_1}{R_3}$, $\frac{R_2}{R_3}$, and $\frac{R_1}{R_2}$ of clusters populating each size group,
and corresponding structural dimensional classifications of $a$-\ce{Te} clusters (see main text and Figure~\num{6} for details).}
\label{tab:gyration_avg}
\begin{tabular}{lcccc}
\toprule
\textbf{Size (atoms)}  & \textbf{$\frac{R_1}{R_3}$} & \textbf{$\frac{R_2}{R_3}$} & \textbf{$\frac{R_1}{R_2}$} & \textbf{Dimensionality} \\
\midrule
1--4   & 4.82~$\pm$~2.17 & 2.71~$\pm$~1.43 & 1.97~$\pm$~0.68 & 1D \\
5--9   & 3.00~$\pm$~0.88 & 1.96~$\pm$~0.55 & 1.69~$\pm$~0.77 & 1D \\
10--19 & 3.82~$\pm$~1.01 & 2.30~$\pm$~0.93 & 1.86~$\pm$~0.71 & 1D \\
20--39 & 2.67~$\pm$~0.89 & 1.62~$\pm$~0.47 & 1.67~$\pm$~0.48 & 2D \\
$>$39  & 2.00~$\pm$~0.41 & 1.62~$\pm$~0.46 & 1.27~$\pm$~0.13 & 3D \\
\bottomrule
\end{tabular}
\end{table}


\clearpage

\providecommand{\latin}[1]{#1}
\makeatletter
\providecommand{\doi}
  {\begingroup\let\do\@makeother\dospecials
  \catcode`\{=1 \catcode`\}=2 \doi@aux}
\providecommand{\doi@aux}[1]{\endgroup\texttt{#1}}
\makeatother
\providecommand*\mcitethebibliography{\thebibliography}
\csname @ifundefined\endcsname{endmcitethebibliography}
  {\let\endmcitethebibliography\endthebibliography}{}

\end{document}